\documentclass[12pt]{article}
\usepackage{epsf}
\makeatletter

\def\newpic#1{}
\voffset= - 0.5cm

\def\hybrid{\topmargin 0pt      \oddsidemargin 0pt
             \headheight 0pt \headsep 0pt

             \textwidth 6.25in       % A4 paper
             \textheight 9.5in       % A4 paper
             \marginparwidth 0.0in
             \parskip 5pt plus 1pt   \jot = 1.5ex}
\catcode`\@=11
\def\marginnote#1{}

\newcount\hour
\newcount\minute
\newtoks\amorpm
\hour=\time\divide\hour by60
\minute=\time{\multiply\hour by60 \global\advance\minute by-\hour}
\edef\standardtime{{\ifnum\hour<12 \global\amorpm={am}%
             \else\global\amorpm={pm}\advance\hour by-12 \fi
             \ifnum\hour=0 \hour=12 \fi
             \number\hour:\ifnum\minute<10 0\fi\number\minute\the\amorpm}}
\edef\militarytime{\number\hour:\ifnum\minute<10 0\fi\number\minute}

\def\draftlabel#1{{\@bsphack\if@filesw {\let\thepage\relax
        \xdef\@gtempa{\write\@auxout{\string
           \newlabel{#1}{{\@currentlabel}{\thepage}}}}}\@gtempa
        \if@nobreak \ifvmode\nobreak\fi\fi\fi\@esphack}
             \gdef\@eqnlabel{#1}}
\def\@eqnlabel{}
\def\@vacuum{}
\def\draftmarginnote#1{\marginpar{\raggedright\scriptsize\tt#1}}

\def\draftlabel#1{{\@bsphack\if@filesw {\let\thepage\relax
        \xdef\@gtempa{\write\@auxout{\string
           \newlabel{#1}{{\@currentlabel}{\thepage}}}}}\@gtempa
        \if@nobreak \ifvmode\nobreak\fi\fi\fi\@esphack}
             \gdef\@eqnlabel{#1}}
\def\@eqnlabel{}
\def\@vacuum{}
\def\draftmarginnote#1{\marginpar{\raggedright\scriptsize\tt#1}}

\def\draft{\oddsidemargin -.5truein
             \def\@oddfoot{\sl preliminary draft \hfil
             \rm\thepage\hfil\sl\today\quad\militarytime}
             \let\@evenfoot\@oddfoot \overfullrule 3pt
             \let\label=\draftlabel
             \let\marginnote=\draftmarginnote
        \def\@eqnnum{(\theequation)\rlap{\kern\marginparsep\tt\@eqnlabel}%
\global\let\@eqnlabel\@vacuum}  }

\def\numberbysection{\@addtoreset{equation}{section}
             \def\theequation{\thesection.\arabic{equation}}}

\def\underline#1{\relax\ifmmode\@@underline#1\else
             $\@@underline{\hbox{#1}}$\relax\fi}

\def\titlepage{\@restonecolfalse\if@twocolumn\@restonecoltrue\onecolumn
          \else \newpage \fi \thispagestyle{empty}\c@page\z@
             \def\thefootnote{\fnsymbol{footnote}} }

\def\endtitlepage{\if@restonecol\twocolumn \else  \fi
             \def\thefootnote{\arabic{footnote}}
             \setcounter{footnote}{0}}  %\c@footnote\z@ }
\relax

\makeatletter
\newdimen\normalarrayskip              % skip between lines
\newdimen\minarrayskip                 % minimal skip between lines
\normalarrayskip\baselineskip
\minarrayskip\jot
\newif\ifold             \oldtrue            \def\new{\oldfalse}
\def\arraymode{\ifold\relax\else\displaystyle\fi} % mode of array entries
\def\eqnumphantom{\phantom{(\theequation)}}     % right phantom in eqnarray
\def\@arrayskip{\ifold\baselineskip\z@\lineskip\z@
         \else
         \baselineskip\minarrayskip\lineskip2\minarrayskip\fi}
\def\@arrayclassz{\ifcase \@lastchclass \@acolampacol \or
\@ampacol \or \or \or \@addamp \or
       \@acolampacol \or \@firstampfalse \@acol \fi
\edef\@preamble{\@preamble
      \ifcase \@chnum
         \hfil$\relax\arraymode\@sharp$\hfil
         \or $\relax\arraymode\@sharp$\hfil
         \or \hfil$\relax\arraymode\@sharp$\fi}}
\def\@array[#1]#2{\setbox\@arstrutbox=\hbox{\vrule
         height\arraystretch \ht\strutbox
         depth\arraystretch \dp\strutbox
         width\z@}\@mkpream{#2}\edef\@preamble{\halign
\noexpand\@halignto
\bgroup \tabskip\z@ \@arstrut \@preamble \tabskip\z@ \cr}%
\let\@startpbox\@@startpbox \let\@endpbox\@@endpbox
      \if #1t\vtop \else \if#1b\vbox \else \vcenter \fi\fi
      \bgroup \let\par\relax
      \let\@sharp##\let\protect\relax
      \@arrayskip\@preamble}
\def\eqnarray{\stepcounter{equation}%
                  \let\@currentlabel=\theequation
                  \global\@eqnswtrue
                  \global\@eqcnt\z@
                  \tabskip\@centering
                  \let\\=\@eqncr
     \halign to \displaywidth\bgroup
        \eqnumphantom\@eqnsel\hskip\@centering
        $\displaystyle \tabskip\z@ {##}$%
        \global\@eqcnt\@ne \hskip 2\arraycolsep
             $\displaystyle\arraymode{##}$\hfil
        \global\@eqcnt\tw@ \hskip 2\arraycolsep
             $\displaystyle\tabskip\z@{##}$\hfil
             \tabskip\@centering
        &{##}\tabskip\z@\cr}
\begingroup\ifx\undefined\newsymbol \else\def\input#1 {\endgroup}\fi
\newfont{\hr}{msbm10}
\newfont{\ams}{msam10}
\def\beq{\begin{equation}}
\def\eeq{\end{equation}}
\def\ba{\beq\new\begin{array}{c}}
\def\ea{\end{array}\eeq}
\def\be{\ba}
\def\ee{\ea}
\def\stackreb#1#2{\mathrel{\mathop{#2}\limits_{#1}}}

\def\Im{{\rm Im}}
\def\Re{{\rm Re}}

\def\d{\partial}
\def\p{\partial}

\def\ha{{1\over 2}}
\def\Bf#1{\mbox{\boldmath $#1$}}

\def\bdelta{{\Bf\delta}}
\def\bomega{{\Bf\omega}}
\def\btau{{\Bf\tau}}
\def\bW{{\bf W}}

\def\bZ{{\bf Z}}
\def\bn{{\bf n}}

\numberbysection
\hybrid

\begin{document}
\setcounter{footnote}{3}
\begin{titlepage}

\title{Integrable Structure of the Dirichlet Boundary Problem in
Multiply-Connected Domains}

\author{I.~Krichever \thanks{Department of Mathematics, Columbia
University, New York, USA, Landau Institute and ITEP, Moscow, Russia}
\and
%\author{
A.~Marshakov \thanks{Max Planck Institute of Mathematics,
Bonn, Germany,
%Theory Department, P.N.
Lebedev Physics Institute
%117924 Moscow, Russia
and ITEP,
%, 117259
Moscow, Russia} %\date{ }
\and A.~Zabrodin
\thanks{Institute of Biochemical Physics
%119991 Moscow, Russia
and ITEP,
%, 117259
Moscow, Russia}}

\date{August 2003}
\maketitle
\vspace{-8cm}

\centerline{
\hfill MPIM -2003 -42}
\centerline{
\hfill ITEP/TH-24/03}
\centerline{
\hfill FIAN/TD-09/03}

\vspace{8cm}

\begin{abstract}

We study the integrable structure of
the Dirichlet boundary problem
in two dimensions and extend the approach
to the case of planar multiply-connected
domains. The solution to the Dirichlet boundary problem
in multiply-connected case is given
through a quasiclassical tau-function,
which generalizes
the tau-function of the dispersionless Toda
hierarchy. It is shown to obey an infinite hierarchy
of Hirota-like equations which directly
follow from properties of the
Dirichlet Green function and from the Fay
identities. The relation to multi-support solutions
of matrix models is briefly discussed.

\end{abstract}

\vfill

\end{titlepage}
\setcounter{footnote}{0}
\section{Introduction}

The Dirichlet boundary problem \cite{C-H}
is to reconstruct a harmonic function in a
bounded domain from its values on the boundary.
Remarkably, this standard problem of complex
analysis, related however to string theory
and matrix models, possesses a hidden integrable
structure \cite{MWZ},
which we clarify further in this paper.
It turns out that variation of a solution
to the Dirichlet problem under variation
of the domain is described by an infinite
hierarchy of non-linear partial differential equations
known (in the simply-connected case) as dispersionless Toda hierarchy.
It is a particular example of the universal hierarchy of
Whitham equations introduced in \cite{KriFun,KriW}.

The quasiclassical tau-function or, more precisely, its logarithm $F$,
is the main new object associated with a family of domains
in the plane. Any domain in the complex plane with sufficiently
smooth boundary can be parameterized by its
moments with respect to a basis of
harmonic functions.
The $F$-function is a function of the full infinite
set of the moments. The first order
derivatives of $F$ are then moments of the complementary
domain. This gives a formal solution to the inverse potential
problem, considered for the simply-connected case in \cite{M-W-Z,W-Z}.
The second order derivatives
are coefficients of the Taylor expansion of the
Dirichlet Green
function and therefore they solve the
Dirichlet boundary problem. These coefficients are constrained by
infinite number of universal (i.e. domain-independent)
relations which, unified in a generating form, just
constitute the dispersionless Hirota equations. For the third order
derivatives (their role in problems of complex analysis
is not yet quite clear) there is a nice ``residue formula''
which allows one to prove \cite{BMRWZ} that $F$ obeys the WDVV
equations.

Below we are going to demonstrate that for planar
multiply-connected domains the
solution to the Dirichlet boundary problem can be performed
in a similar way. Specifically, we consider domains
which are obtained by cutting several ``holes''
in the complex plane.
Boundaries of the holes are assumed to be smooth simple
non-intersecting curves.
In this case, the complete set of independent
variables can be again identified with the set of
harmonic moments. However, a choice of the {\it proper} basis of
harmonic functions in a multiply-connected domain becomes crucial for our
approach. It turns out that the Laurent polynomials which were used
in the simply-connected case should be replaced by the basis analogous
to the one introduced in \cite{kn} -- a ``global" generalization of
the Laurent basis for
algebraic curves of arbitrary genus. The basis
has to be also enlarged to include harmonic functions
with multi-valued analytic part.
This results in an additional finite set of extra variables.
We construct the $F$-function and prove that
its second derivatives
satisfy non-linear relations, which generalize the Hirota
equations of the dispersionless Toda hierarchy.
These relations are derived
from the Fay identities \cite{Fay} for the Riemann theta functions
on the Jacobian of Riemann surface obtained as the {\em Schottky double}
of the plane with given holes.

We note that extra variables,
specific for the multiply-connected case,
can be chosen in different ways and possess
different geometric interpretations,
depending on the choice of basis of
homologically non-trivial cycles on the
Schottky double.
The corresponding $F$-functions are shown to be
connected by a duality transformation -- a (partial) Legendre transform,
with the generalized Hirota relations being the same.

Now let us give a bit more expanded description of the
Dirichlet problem in planar domains.
Let ${\sf D^c}$ be a domain in the complex plane
bounded by one or several non-intersecting smooth curves.
It will be convenient to realize ${\sf D^c}$
as a complement to another domain ${\sf D}$, having
one or more connected components, and
to consider the Dirichlet problem in ${\sf D^c}$:
to find a harmonic
function $u(z)$ in ${\sf D^c}$ such that it is continuous
up to the boundary, $\d {\sf D^c}$, and equals a given function
$u_0(\xi )$ on the boundary.
The problem has
a unique solution written in terms of the Dirichlet
Green function $G(z,\xi )$:
\be\label{Dirih}
u(z)=
- \frac{1}{2\pi}\oint_{\d {\sf D^c}}
u_0(\xi )\p_{n} G(z,\xi ) |d\xi |
\ee
where $\p_n$ is the normal derivative on the boundary
with respect to the second variable,
the normal vector $\vec n$ is directed
inward ${\sf D^c}$, and $|d\xi| := dl( \xi)$ is an infinitesimal
element of the length of the boundary $\p {\sf D^c}$.

The main object to study is, therefore, the Dirichlet Green function.
It is uniquely determined by the following properties \cite{C-H}:
\begin{itemize}
\item[($G1$)] The function
$G(z,z')$
is symmetric and harmonic everywhere in ${\sf D^c}$
(including $\infty$ if ${\sf D^c} \ni \infty$) in
both arguments except $z=z'$ where
$G(z,z')=\log |z-z'| +\ldots $ as $z\to z'$;
\item[($G2$)] $G(z,z')=0$ if any one
of the variables $z$, $z'$ belongs to the boundary $\p {\sf D^c}$.
\end{itemize}
Note that the definition implies that $G(z,z')<0$
inside ${\sf D^c}$. In particular, $\p_n G(z, \xi)$ is
strictly negative for all $\xi \in \p {\sf D^c}$.

If ${\sf D^c}$ is simply-connected (note that we assume $\infty\in{\sf
D^c}$), i.e., the boundary has only
one component,
the Dirichlet problem is equivalent to finding
a bijective conformal map from
${\sf D^c}$ onto the complement to unit disk
or any other reference domain for which the Green function
is known explicitly. Such bijective conformal map $w(z)$ exists by virtue
of the Riemann mapping theorem, then
\be\label{Gconf}
G(z, z')=\log \left |
\frac{w(z)-w(z')}{w(z)\overline{w(z')} -1} \right |
\ee
where bar means complex conjugation.
It connects the Green function at two points
with the conformal map normalized at some third point
(say at $z=\infty$: $w(\infty )=\infty$).
It is this formula which allows one to derive the Hirota
equations for the tau-function of the Dirichlet problem in the most
economic and transparent way \cite{MWZ} (see also sect.~\ref{ss:simply} below).

For multiply-connected domains, formulas of this type
based on conformal maps
do not really exist. In general, there is no canonical
choice of the reference domain, moreover, the shape of a
reference domain depends on ${\sf D^c}$ itself.
In fact, as we demonstrate in the paper,
the correct extension of (\ref{Gconf})
needed for derivation of the generalized Hirota equations
follows from a different direction which is no longer explicitly related
to bijective conformal maps. Namely, logarithm of the conformal map
$\log w(z)$ should be replaced now by the Abel
map from the Schottky double of ${\sf D^c}$
to the Jacobi variety of this Riemann surface, and
the rational function under the logarithm in (\ref{Gconf})
is substituted
by ratio of the prime forms or Riemann theta-functions.

\begin{figure}[tb]
\epsfysize=5cm
\centerline{\epsfbox{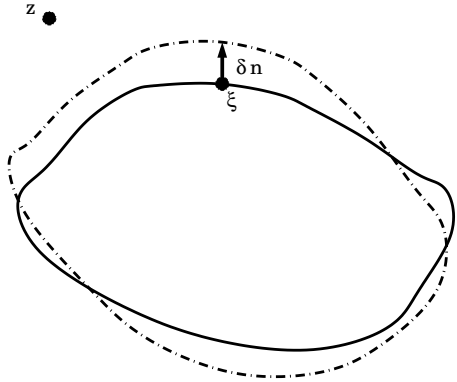}}
\caption{\sl A ``pictorial" derivation of the
Hadamard formula. We consider
a small deformation of the domain, with the new boundary being
depicted by the dashed line. According to ($G2$) the
Dirichlet Green function vanishes, $G(z,\xi)=0$,
if $\xi$ belongs to the old boundary. Then
the variation $\delta G(z,\xi)$ is simply equal to
the new value, i.e. in the leading order
$\delta G(z,\xi) = -\delta n(\xi)\d_n G(z,\xi)$.
Now notice that $\delta G(z, \xi)$ is a {\it harmonic
function} (the logarithmic singularity cancels since it is
the same for both old and new functions)
with the boundary value
$-\delta n(\xi)\d_n G(z,\xi)$.
Applying (\ref{Dirih})
one obtains (\ref{Hadam}). The argument
is the same for both simply-connected and
multiply-connected domains.}
\label{fi:hadamar}
\end{figure}

We show that the Green function
of multiply-connected domains admits a representation
through the logarithm of the tau-function of the form
\be\label{GGG}
G(z,z')=\log \left |\frac{1}{z}-\frac{1}{z'}\right |
+\frac{1}{2}\nabla (z)\nabla (z')F
\ee
Here $\nabla (z)$
%(see (\ref{}) below)
is certain vector field on the moduli space of boundary curves,
therefore it can be represented as a (first-order) differential
operator w.r.t. harmonic moments with constant (in moduli) coefficients depending,
however, on the point $z$
as a parameter.

In this paper we also obtain
similar formulas for the harmonic measures of
the boundary components and for the Abel map.
A combination of these formulas with the Fay identities
yields the generalized
Hirota-like equations for the tau-function $F$.

Our main tool
is the Hadamard variational formula \cite{Hadamard} which gives
variation of the Dirichlet Green function under small deformations
of the domain in terms of the Green function itself:
\be\label{Hadam}
\delta G(z, z')=\frac{1}{2\pi}\oint_{\d {\sf D^c}}
\p_{n}G(z, \xi)\p_{n}G(z', \xi)\delta n(\xi)|d\xi |.
\ee
Here $\delta n(\xi)$ is the normal displacement (with sign!)
of the boundary under the deformation,
counted along the normal
vector at the boundary point $\xi$.
It was shown in \cite{MWZ} that this remarkable formula
is a key to all
integrable properties of the Dirichlet problem. An
extremely simple ``pictorial" derivation of the formula
(\ref{Hadam}) is
presented in fig.~\ref{fi:hadamar}.

We start with a brief recollection of the
results for the simply-connected case in sect.~\ref{ss:simply}.
However, instead of ``bump'' deformations used
in \cite{MWZ} we work here with their rigorously defined versions --
a family of infinitesimal deformations
which we call elementary ones.
This approach is basically motivated by the theory of
interface dynamics in viscous
fluids, which is known to be closely connected with the
formalism developed in \cite{MWZ} and
in the present paper (see \cite{M-W-Z} for details).

In sect.~\ref{ss:taumu}
we introduce local
coordinates in the space
of planar multiply-connected domains
and express the elementary deformations
in these coordinates.
Using the Hadamard formula, we then observe
remarkable symmetry or ``zero-curvature'' relations
which connect elementary deformations of the Green
function and harmonic measures. The
existence of the tau-function and the formula
(\ref{GGG}) for the Green function directly
follow from these relations.
In sect.~\ref{ss:dual} we make a Legendre transform to
another set of local coordinates in the space
of algebraic multiply-connected domains, which is
in a sense dual to the original one. In these
coordinates, eq.~(\ref{GGG}) gives another
version of the Green function which solves
the so-called modified Dirichlet problem.
We also discuss the relation to
multi-support solutions of matrix models
in the planar large $N$ limit.

In sect.~5 we combine the results outlined
above with the representation of the Green
function in terms of the prime form on the
Schottky double. This allows us to obtain
an infinite system of partial differential equations
on the tau-function which generalize the
dispersionless Hirota equations.

\section{The Dirichlet problem for simply-connected
domains and dispersionless Hirota equations
\label{ss:simply}}

In this section we rederive the results from
\cite{MWZ} for the
simply-connected case in a slightly different manner,
more suitable for further generalizations. At the same
time we show that the results of \cite{MWZ} obtained for
{\it analytic} curves can be easily extended to the {\it smooth} case.

Let ${\sf D}$ be a connected domain in the complex plane
bounded by a simple
smooth curve.
We consider the exterior Dirichlet problem in
${\sf D^c}={\bf C}\setminus {\sf D}$ which is the complement of
${\sf D}$ in the whole (extended) complex plane.
Without loss of generality,
we assume that ${\sf D}$ is compact and
contains the point $z=0$.
Then ${\sf D^c}$ is a simply-connected domain
on the Riemann sphere containing $\infty$.

\paragraph{Harmonic moments and deformations of the boundary.}

Let $t_k$ be moments of the domain
${\sf D^c}={\bf C}\setminus {\sf D}$
defined with respect
to the harmonic functions
$\{ {z^{-k}/ k}\}$, $k>0$:
\be\label{momt}
t_k= \, - \, \frac{1}{\pi k}
\int_{{\sf D^c}}z^{-k} \,d^2 z\,, \,\;\;\;\;\;k=1,2,\ldots
\ee
and $\{{\bar t}_k\}$ be the complex conjugate moments, i.e.
$ {\bar t}_k = -{1\over\pi k}\int_{{\sf D^c}}d^2 z
\bar z^{-k}$.
The Stokes formula
represents the harmonic moments as contour
integrals
\beq\label{momtcont}
t_{k}=\frac{1}{2\pi i k}\oint_{\d {\sf D}}
z^{-k} \bar z dz
\eeq
providing, in particular, a regularization of possibly divergent
integrals (\ref{momt}).
Besides, we denote by $t_0$
the area (divided by $\pi$) of the domain ${\sf D}$:
\be\label{t0}
t_0 =\frac{1}{\pi}\int_{{\sf D}} d^2z
\ee
The harmonic moments of ${\sf D^c}$ are coefficients of
the Taylor expansion of the potential
\be\label{dd1}
\Phi (z) =-\frac{2}{\pi}\int_{{\sf D}} \log |z-z'|d^2z'
\ee
induced by the domain ${\sf D}$ filled by two-dimensional
Coulomb charges with the uniform density $\rho = -1$.
Clearly, $\d_z \d_{\bar z}\Phi (z) =-1$ if $z\in {\sf D}$ and vanishes
otherwise, so around the origin (recall that ${\sf D}\ni 0$)
the potential equals to $-|z|^2$ plus a harmonic function, i.e.
\be\label{deftk}
\Phi (z)-\Phi (0)=-|z|^2 +\sum_{k\geq 1}
\left (t_k z^k +\bar t_k \bar z^k \right )
\ee
and one can verify that $t_k$ are just
given by (\ref{momt}).

For analytic boundary curves, one may introduce
the Schwarz function associated with the curve.
The function
$$
\p_z \Phi (z) =-\frac{1}{\pi}\int_{{\sf D}}
\frac{d^2 z'}{z-z'}
$$
is continuous across the boundary and holomorphic for
$z\in {\sf D^c}$ while for $z \in {\sf D}$ the function
$\p_z \Phi +\bar z$ is holomorphic.
If the boundary is an analytic curve,
both these functions can be analytically continued outside
the regions where they were originally defined, and, therefore,
there exists a function, $S(z)$,
analytic in some strip-like neighborhood
of the boundary contour, such that $S(z)=\bar z$ on the contour.
In other words, $S(z)$ is the analytic continuation of
$\bar z$ away from the boundary contour, this function
completely determines the shape of the boundary and is called
{\it the Schwarz function} \cite{Davis}.
In general we are going to work with smooth
curves, not necessarily analytic, when the
Schwarz function does not exist as an analytic function.
Nevertheless, it appears to be useful below to define the class
of boundary contours with nice algebro-geometric
properties.

The basic fact of the theory of deformations of
closed smooth curves is that the (in general
complex) moments $\{ t_k, \bar t_k\}\equiv \{ t_{\pm k}\}$
supplemented by the real variable
$t_0$ form a set of local coordinates in
the ``moduli space" of smooth closed curves
\cite{Kriunp} (see also \cite{T}).

\medskip
\noindent
{\it Important remark.}
This means that: (a) under any small deformation of
the domain the set ${\bf t} = \{t_0, t_{\pm k}\}$
is subject to a small change;
(b) on the space of smooth closed curves there
exist vector fields $\p_{t_k}$ such that
$\p_{t_k}t_n=\delta_{kn}$,
which are represented in terms of infinitesimal
normal displacements of the boundary that
change $x_k={\rm Re} \, t_k$ or $y_k={\rm Im} \, t_k$
keeping all the other moments
fixed; (c) the corresponding infinitesimal
displacements can be locally integrated.
The latter means that for each domain ${\sf D^c}$
with moments $\{t_0,t_{\pm k}\}$ and for
an arbitrary integer $N$ there exist
constants $\epsilon _m$,  $|m|\leq N$,
such that for any set
$\{t'_0,t'_{\pm k}\}$ with $|t'_m-t_m|<\epsilon_m$,
$m\leq N$, $t_m'=t_m$, $|m|>N$,
in the neighborhood of ${\sf D^c}$
there is a unique domain
with the moments $\{t'_0,t'_{\pm k}\}$. We adopt this restricted
notion of the local coordinates throughout the paper.
It would be very interesting to find
conditions on the infinite sets $\epsilon_k$
for the corresponding rectangles
to form an open set in an infinite-dimensional
variety of smooth curves.
We plane to address this problem elsewhere.

Let us present a proof of this statement which later will be easily adjusted
to the case of multiply-connected domains. At the same time this proof
allows one to derive a deformation of the domain with respect to the variables
$t_k$. Suppose there is a one-parametric deformation ${\sf D}(t)$ (with
some real parameter $t$) of ${\sf D}={\sf D}(0)$ such that all $t_k$
are preserved: $\p_t t_k =0$, $k\geq 0$. Let us prove that
such a deformation is trivial. The proof is based on two key observations:

\begin{itemize}
\item
{\it The difference of the boundary values
$\p_tC^{\pm}(\zeta)d\zeta$
of the derivative of the Cauchy
integral}
\be\label{C}
C(z)dz=
{dz\over 2\pi i}\oint_{\p {\sf D}} {\bar \zeta d\zeta
\over \zeta -z}
\ee
{\it is purely imaginary differential on the boundary of ${\sf D}$}.

Indeed, let $\zeta(\sigma ,t)$ be a
parameterization of the curve $\p {\sf D}(t)$. Denote the value of
the differential (\ref{C})
by $C^-(z)dz$ for $z\in {\sf D^c}$
and by $C^+(z)dz$ for $z\in {\sf D}$.
Taking the $t$-derivative of (\ref{C}) and integrating by parts one gets
\be\label{C1}
\p_tC(z)dz=
{dz\over 2\pi i}\oint_{\p {\sf D}} \left({\bar \zeta_t \zeta_{\sigma}+
\bar \zeta \zeta_{t,\,\sigma}
\over \zeta -z}-{\bar \zeta \zeta_{\sigma}\zeta_t
\over (\zeta -z)^2}\right)d\sigma =
\\
={dz\over 2\pi i}\oint_{\p {\sf D}} \left({\bar \zeta_t \zeta_{\sigma}-
\bar \zeta_{\sigma} \zeta_{t}
\over \zeta -z}\right)d\sigma
\ee
Hence,
$$
\left(\p_tC^{+}(\zeta)-\p_tC^{-}(\zeta)\right)d\zeta
=\p_t\bar \zeta d\zeta -\p_t\zeta d\bar \zeta =
2i \Im\left(\p_t\bar \zeta d\zeta \right)
$$
is indeed purely imaginary.

\bigskip

\item
{\it If a $t$-deformation preserves all the
moments $t_k$, $k\geq 0$, the differential\\
$\p_t \bar\zeta d\zeta-\p_t\zeta d\bar\zeta$ extends to a holomorphic
differential in ${\sf D^c}$}.

If $|z|<|\zeta |$ for all $ \zeta \in \p {\sf D}$,
then we can expand:
\beq\label{7}
\p_tC^+(z)dz=\frac{\p}{\p t}
\left ( {dz\over 2\pi i}\sum_{k=0}^{\infty}
z^k \oint_{\p {\sf D}} \zeta^{-k-1}\bar \zeta d\zeta \right ) =
\sum_{k=1}^{\infty} k\left(\p_t t_{k}\right)z^{k-1}dz=0
\eeq
and, since $C^+$ is analytic in ${\sf D}$,
we conclude that $\p_t C^+ \equiv 0$.
The expression
$\p_t \bar\zeta d\zeta-\p_t\zeta d\bar\zeta$ is the boundary value
of the differential
$- \p_t C^-(z)dz$ which has at most simple pole at the infinity
and holomorphic everywhere else in ${\sf D^c}$. The equality
$$
\p_t t_0 =\frac{1}{2\pi i}\oint_{\p {\sf D}}
(\p_t \bar\zeta d\zeta  - \p_t \zeta d\bar \zeta ) = 0
$$
then implies that the
residue at $z=\infty$ vanishes, therefore $\p_t C^-(z)dz$ is holomorphic.
\end{itemize}

Any holomorphic differential which is purely imaginary
along the boundary of a simply-connected
domain must be zero in this domain.
Indeed, the real part of the harmonic continuation
of the integral of this
differential is a harmonic function with a constant
boundary value. Such a function must be constant
by virtue of the uniqueness of the solution to the
Dirichlet problem.
Another proof
relies on the Schwarz symmetry principle
and the standard Schottky double construction
(see the next section for details).
Consider the compact Riemann surface obtained by attaching to
${\sf D^c}$ its
complex conjugated copy along the boundary.
Since $\p_t C^-dz$
is imaginary along the boundary,
we conclude, from the Schwarz symmetry principle,
that $\p_tC^-dz$ extends to a
globally defined holomorphic differential on this
compact Riemann surface, which has genus zero.
Therefore, such a differential
is equal to zero. Hence we conclude that
$\p_t \bar\zeta d\zeta-\p_t\zeta d\bar\zeta=0$.
This means that
the vector $\p_t \zeta$ is tangent to the boundary.
Without loss of generality we can always assume that a parameterization of
$\p{\sf D}(t)$ is chosen so that
$\p_t \zeta (\sigma , t)$ is normal to the
boundary.
Thus, the $t$-deformation of the boundary preserving all harmonic moments is
trivial.

The fact that the set of harmonic moments is not
overcomplete follows from the explicit construction
of vector fields in the space of domains that changes
any harmonic moment keeping all the others fixed
(see below).

\paragraph{Elementary deformations and the operator
$\nabla (z)$.}

\begin{figure}[tb]
\epsfysize=5cm
\centerline{\epsfbox{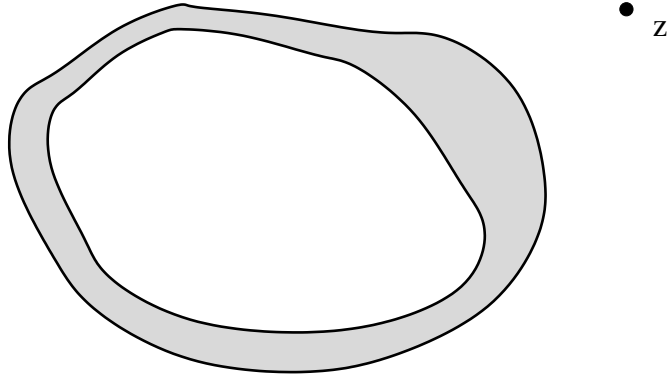}}
\caption{\sl The elementary deformation
with the base point $z$.}
\label{fi:bump}
\end{figure}

Fix a point $z\in {\sf D^c}$ and consider
a special infinitesimal
deformation of the domain such that the normal displacement
of the boundary is proportional to the gradient of the
Green function $G(z,\xi)$ at the boundary point
(fig.~\ref{fi:bump}):
\be\label{small}
\delta n(\xi)
=-\frac{\epsilon}{2}\p_n G(z, \xi)
\ee
For any sufficiently smooth initial boundary
this deformation is well-defined as $\epsilon \to 0$.
We call infinitesimal deformations from this family,
parametrized by $z\in {\sf D^c}$, the
{\it elementary deformations}. The point $z$ is
refered to as the
{\it base point} of the deformation.
Note that
since $\p_n G <0$ (see the remark after the definition of the
Green function
in the Introduction), $\delta n $
for the elementary deformations is either strictly positive
or strictly negative depending of the sign of $\epsilon$.

Let $\delta_z$ be variation of any quantity under the
elementary deformation with the base point $z$.
It is easy to see that
$\delta_z t_0 =\epsilon$,
$\delta_z t_k =
\epsilon z^{-k}/k$.
Indeed,
\be
\label{spe}
\begin{array}{l}
\displaystyle{\delta_z t_0 =
\frac{1}{\pi}\oint
\delta n(\xi) |d\xi|=
-\frac{\epsilon}{2\pi }\oint
\p_n G(z, \xi ) |d\xi|= \epsilon}
\\ \\
\displaystyle{\delta_z t_k =
\frac{1}{\pi k}\oint \xi^{-k}
\delta n(\xi) |d\xi|=
-\frac{\epsilon}{2\pi k}\oint \xi^{-k}
\p_n G(z, \xi ) |d\xi|= {\epsilon \over k} z^{-k}}
\end{array}
\ee
by virtue of the Dirichlet formula (\ref{Dirih}).
Note that the elementary deformation with the base
point at $\infty$ keeps all moments except $t_0$ fixed.
Therefore, the deformation which changes
only $t_0$ is given
by $\delta n(\xi )=-\frac{\epsilon}{2} \p_n G(\infty , \xi )$.

Now we can explicitly define the deformations that
change only either $x_k =\mbox{Re}\, t_k$ or
$y_k = \mbox{Im}\, t_k$
keeping all other moments fixed.
As is clear from (\ref{spe}),
the corresponding $\delta n(\xi)$
is given by the real or imaginary part of
normal derivative
of the function
\beq\label{Hka}
H_k (\xi )=\frac{1}{2\pi i}
\oint_{\infty} z^k \p_z  G(z,\xi )dz
\eeq
at the boundary.
Here the contour integral goes around infinity.
Namely, the normal displacements
$\delta n(\xi )=\epsilon \, {\rm Re}\, (\p_n H_k (\xi ))$
and
$\delta n(\xi )=\epsilon \, {\rm Im}\, (\p_n H_k (\xi ))$
change the real and imaginary part of $t_k$ by
$\pm \epsilon$ respectively keeping all other moments
fixed.

These deformations allow one to introduce the vector fields
$$
\frac{\p}{\p t_0}\,,
\;\;\;\;
\frac{\p}{\p x_k}\,,
\;\;\;\;
\frac{\p}{\p y_k}
$$
in the space of domains which are locally well-defined.
Existence of such vector fields
means that the variables $t_k$ are independent.
For $k>0$ it is more convenient to use their linear combinations
$$
\frac{\p}{\p t_k}=\frac{1}{2} \left (
\frac{\p}{\p x_k}  -  i\,
\frac{\p}{\p y_k} \right ),
\;\;\;\;\;\
\frac{\p}{\p \bar t_k}=\frac{1}{2} \left (
\frac{\p}{\p x_k}  + i\,
\frac{\p}{\p y_k} \right )
$$
which span the complexified tangent space to the space
of simply-connected domains (with fixed area $t_0$).
If $X$ is any functional of our domain
locally representable as a function
of harmonic moments, $X=X({\bf t})$,
the vector fields $\p_{t_0}$,
$\p_{t_k}$, $\p_{\bar t_k}$ can be understood
as partial derivatives acting to the function
$X({\bf t})$.

Consider the variation $\delta_z X$ of
a functional $X=X({\bf t})$ under the
elementary deformation with the base point $z$.
In the leading order in $\epsilon$ we have:
\be
\label{D4}
\delta_z X =
\sum_k \frac{\p X}{\p t_k}\, \delta_z t_k =
\epsilon \nabla (z)X
\ee
where the differential operator $\nabla (z)$
is given by
\be\label{D2}
\nabla (z)
=\p_{t_0} +\sum_{k\geq 1} \left (
\frac{z^{-k}}{k} \p_{t_k} +
\frac{\bar z^{-k}}{k}\p_{\bar t_k}\right )
\ee
The right hand side suggests that
for functionals $X$ such that the series
$\nabla(z) X$ converges everywhere in ${\sf D^c}$ up to the
boundary $\delta_z X$ is a harmonic function
of the base point $z$.

Note that in \cite{MWZ} we have used the ``bump" deformation
and continued it harmonically into ${\sf D^c}$.
In fact, it was the
elementary deformation (\ref{spe})
$\delta_z \propto \oint |d\xi| \p_n G(z,\xi)
\delta^{\rm bump}(\xi)$ that was
really used. The ``bump" deformation
should be understood as
a (carefully taken) limit of $\delta_z$ when the point $z$ tends
to the boundary $\p{\sf D^c}$.

\paragraph{The Hadamard formula as integrability
condition.}

Variation of the Green function under
small deformations of the domain is
known due to Hadamard, see eq.\,(\ref{Hadam}).
To find how the Green function changes under small
variations of the harmonic moments, we fix
three points $a,b,c \in {\bf C}\setminus {\sf D}$ and
compute $\delta_{c}G(a,b)$ by means of the
Hadamard formula (\ref{Hadam}).
Using (\ref{D4}), one can identify the result
with the action of the vector field $\nabla (c)$
on the Green function:
\be\label{Th0}
\nabla (c)G(a,b) =
-\, \frac{1}{4\pi }\oint_{\d {\sf D}}
\p_n G(a , \xi) \p_n G(b , \xi)
\p_{n} G(c , \xi) |d\xi|
\ee
Remarkably, the r.h.s. of (\ref{Th0})
is {\it symmetric} in all three arguments, i.e.
\be\label{symha}
\nabla  (a)G(b,c)
=\nabla (b)G(c,a)
=\nabla (c)G(a,b)
\ee
This is the key relation which allows one
to represent the Dirichlet problem
as an integrable hierarchy of non-linear
differential equations \cite{MWZ}, (\ref{symha}) being
the integrability condition of the hierarchy.

It follows from (\ref{symha}) (see \cite{MWZ}
for details) that there exists a function
$F=F({\bf t})$ such that
\be
\label{gf}
G(z,z') = \log\left|{1\over z} - {1\over z'}\right|
+ \ha \nabla (z)\nabla (z')F
\ee
We note that
existence of such a representation
of the Green function was first conjectured
by Takh\-ta\-jan.
For the simply-connected case,
this formula was obtained in \cite{K-K-MW-W-Z}
(see also \cite{T} for a detailed proof
and discussion).
The function $F$
is (logarithm of) the tau-function of
the integrable hierarchy. In \cite{K-K-MW-W-Z}
it was called the tau-function of the (real analytic) curves
-- the boundary contours $\p{\sf D}$ or $\p{\sf D^c}$.

\paragraph{Dispersionless Hirota equations.}
Combining (\ref{gf}) and (\ref{Gconf}), we obtain
the relation
\beq\label{Gconf1}
\log \left |
\frac{w(z)-w(z')}{w(z)\overline{w(z')} -1} \right |^2
=\log\left|{1\over z} - {1\over z'}\right|^2
+ \nabla (z)\nabla (z')F
\eeq
which implies
an infinite hierarchy of differential equations
on the function $F$.
It is convenient to normalize
the conformal map $w(z)$ by the conditions that
$w(\infty )=\infty$ and $\p_z w(\infty )$ is real,
so that
\beq\label{confrad}
w(z)=\frac{z}{r}+O(1) \;\;\;
\mbox{as $z\to \infty$}
\eeq
where the real number
$r=\lim_{z\to\infty} {dz/ dw(z)}$
is called the
(external) conformal radius of the domain
${\sf D}$ (equivalently, it can be defined through
the Green function as
$\log r = \lim_{z\to \infty}(G(z, \infty)+\log |z|)$,
see \cite{Hille}).
Then, tending $z' \to\infty$ in
(\ref{Gconf1}),
one gets
\be\label{sec3a}
\log |w(z)|^2=\log |z|^2 - \p_{t_0}\nabla (z)F
\ee
The limit $z\to \infty$ of this equality
yields a simple formula for the conformal radius:
\be\label{sec7}
\log r^2 = \p_{t_{0}}^2 F
\ee

Let us now separate holomorphic and
antiholomorphic parts of these equations,
introducing the
holomorphic and antiholomorphic parts
of the operator $\nabla (z)$ (\ref{D2}):
\beq\label{Dhol}
D(z)=\sum_{k\geq 1}\frac{z^{-k}}{k}
\p_{t_k}\,,
\;\;\;\;\;
\bar D(\bar z)=\sum_{k\geq 1}\frac{\bar z^{-k}}{k}
\p_{\bar t_k}\,,
\eeq
Rewrite
(\ref{Gconf1}) in the form
$$
\begin{array}{ll}
&\displaystyle{
\log \left (
\frac{w(z)-w(z')}{w(z)\overline{w(z')} -1} \right )
-\log \left (\frac{1}{z}-\frac{1}{z'}\right )
-\left (\frac{1}{2}\p_{t_0} +D(z)\right )
\nabla (z' ) F }
\\&\\
=&\displaystyle{
-\log \left (
\frac{\overline{w(z)}-
\overline{w(z')}}{w(z')\overline{w(z)} -1} \right )
+\log \left ( \frac{1}{\bar z}- \frac{1}{\bar z'}\right )
+\left (\frac{1}{2}\p_{t_0} +\bar D(\bar z)\right )
\nabla (z') F }
\end{array}
$$
The l.h.s. is a holomorphic function of $z$ while
the r.h.s. is antiholomorphic. Therefore, both are equal to
a $z$-independent term which can be found from the
limit $z\to \infty$.
As a result, we obtain the equation
\be\label{sec6}
\log \left (
\frac{w(z)-w(z')}{w(z)- (\overline{w(z')})^{-1} } \right )
=\log \left ( 1-\frac{z'}{z}\right )
+D(z)\nabla (z') F
\ee
which, as $z' \to \infty$, turns into the formula
for the conformal map $w(z)$:
\be\label{sec4}
\log w(z)=\log z
-\frac{1}{2}\p^{2}_{t_0}F -\p_{t_0}D(z)F
\ee
(here we also used (\ref{sec7})).
Proceeding in a similar way, one can rearrange (\ref{sec6})
in order to write it separately
for holomorphic and antiholomorphic parts in $z'$:
\be\label{sec5}
\log \frac{w(z)-w(z')}{z-z'}\, =\,
- \,\frac{1}{2}\p_{t_0}^2\, F +
D(z) D(z') F
\ee
\be\label{511}
-\, \log \left (1- \frac{1}{w(z)
\overline{w(z')}}\right )=
D(z)\bar D(\bar z' )F
\ee

Writing down eqs.\,(\ref{sec5})
for the pairs of points $(a,b)$, $(b,c)$ and $(c,a)$ and
summing up the exponentials of the both sides of each equation
one arrives at the relation
\be\label{Hir1}
(a-b)e^{D(a)D(b)F}
+(b-c)e^{D(b)D(c)F}
+(c-a)e^{D(c)D(a)F} =0
\ee
which is the dispersionless Hirota equation (for the KP
part of the two-dimensional Toda lattice hierarchy)
written in the symmetric form.
This equation can be regarded
as a very degenerate case of the trisecant Fay
identity \cite{Fay}.
It encodes the algebraic relations between the second
order derivatives
of the function $F$. As $c \to \infty$, we get
these relations
in a more explicit but less symmetric form:
\be\label{Hir2}
1-e^{D(a)D(b )F}=
\frac{D(a)-D(b )}{a-b}\,\p_{t_1}F
\ee
which makes it clear that the totality of
second derivatives
$F_{ij}:=\p_{t_i}\p_{t_j}F$
are expressed through the derivatives
with one of the indices put equal to unity.

More general equations of the dispersionless Toda hierarchy
obtained in a similar
way by combining eqs.\,(\ref{sec4}), (\ref{sec5}) and
(\ref{511})
include derivatives w.r.t. $t_0$
and $\bar t_k$:
\be\label{Hir3} (a-b)e^{D(a)D(b )F}
=ae^{-\p_{t_0}D(a)F}
-b e^{-\p_{t_0}D(b)F}
\ee
\be\label{Hir4}
1-e^{-D(z)\bar D(\bar z )F}=\frac{1}{z\bar z}
e^{ \p_{t_0} \nabla (z) F}
\ee
These equations allow one to express the second derivatives
$\p_{t_m}\p_{t_n}F$,
$\p_{t_m}\p_{\bar t_n}F$ with $m,n\geq 1$
through the derivatives
$\p_{t_0}\p_{t_k}F$,
$\p_{t_0}\p_{\bar t_k}F$.
In particular, the dispersionless Toda equation,
\be\label{Toda}
\p_{t_1}\p_{\bar t_1}F =e^{\p_{t_0}^{2}F}
\ee
which follows from (\ref{Hir4}) as $z \to \infty$,
expresses $\p_{t_1}\p_{\bar t_1}F$ through $\p_{t_0}^{2}F$.

For a comprehensive exposition
of Hirota equations for dispersionless
KP and Toda hierarchies we refer the reader to
\cite{C-K,T-T}.

\paragraph{Integral representation of the tau-function.}

Eq.\,(\ref{gf}) allows one to obtain a representation
of the tau-function as a double integral over the domain ${\sf D}$.
Set $\tilde \Phi (z):=\nabla (z)F$.
One is able to determine this function
via its variation under the elementary
deformation:
\beq\label{dual1a}
\delta_a \tilde \Phi (z) =
-2\epsilon \log \left |a^{-1}-z^{-1}\right | +
2\epsilon G(a,z)
\eeq
which is read from eq.\,(\ref{gf}) by virtue of
(\ref{D4}).
This allows one to identify $\tilde \Phi$
with the ``modified potential"
$\tilde \Phi (z)=\Phi (z)-\Phi (0) +t_0 \log |z|^2$,
where $\Phi$ is given by (\ref{dd1}).
Thus we can write
\be
\label{modpot}
\nabla (z)F =\tilde \Phi (z)=
-\frac{2}{\pi} \int_{{\sf D}} \log |z^{-1}-\zeta^{-1}| d^2 \zeta
= v_0 + 2\Re\sum_{k>0}\frac{v_k}{k}z^{-k}
\ee
The last equality is to be understood as
the Taylor expansion around infinity.
The coefficients $v_k$ are
moments of the interior domain
(the ``dual'' harmonic moments) defined as
\be
\label{vk}
v_k= \frac{1}{\pi }\int_{{\sf D}}z^{k}\,d^2 z \ \ \ (k>0)\,,
\;\;\;\;\;
v_0 =-\Phi(0)=\frac{2}{\pi}\int_{{\sf D}} \log |z|d^2 z
\ee                      From (\ref{modpot}) it is clear that
\be\label{vk1}
v_k =\p_{t_k}F\,,
\;\;\;\;k\geq 0
\ee
i.e., the moments of the complementary domain ${\sf D}$
(the ``dual'' moments) are
completely determined by the function $F$ of
harmonic moments of ${\sf D^c}$.

In a similar manner, one
arrives at the integral representation of the tau-function. Comparing
(\ref{modpot}) with (\ref{dual1a}) one can easily notice that the meaning
of the elementary deformation $\delta_\xi$ or the operator $\nabla(\xi)$
formally applied at the boundary point $\xi\in\p {\sf D}$
(where $G(z,\xi)=0$) is attaching
a ``small piece" to the integral over the domain ${\sf D}$ (the ``bump" operator
from \cite{MWZ}). Using this fact and interpreting (\ref{modpot}) as a
variation $\delta_z F$ we arrive at the double-integral representation
of the tau-function
\be\label{F}
F=-\frac{1}{\pi^2}\int_{{\sf D}} \! \int_{{\sf D}}
\log |z^{-1} -\zeta^{-1}| d^2 z d^2 \zeta
\ee
or
\be\label{F0}
F=\frac{1}{2\pi}\int_{{\sf D}}
\tilde \Phi (z) d^2 z =\frac{1}{2\pi}\int_{\sf D}
\left(\Phi (z) -2\Phi (0)\right) d^2 z
\ee
As we see below,
the main formulas from this paragraph remain
intact in the multiply-connected case.

\section{The Dirichlet problem and the tau-function in
the multiply-connected case
\label{ss:taumu}}

\begin{figure}[tb]
%\hspace*{2cm}
\epsfysize=7.5cm
\centerline{\epsfbox{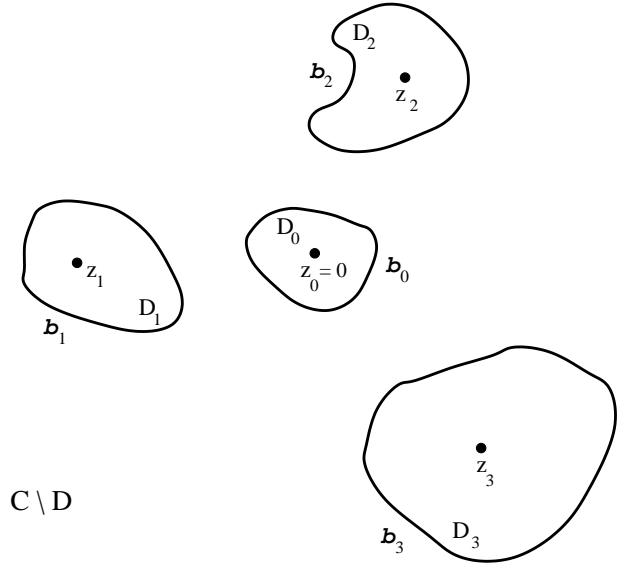}}
\caption{\sl A multiply-connected domain
${\sf D^c}={\bf C}\setminus {\sf D}$ for $g=3$.
The domain ${\sf D} = \bigcup_{\alpha=0}^3 {\sf D}_\alpha$
consists of $g+1=4$ disconnected
parts ${\sf D}_\alpha$ with the boundaries ${\sf b}_\alpha$.
To define the complete set of harmonic
moments, we also need the auxiliary points
$z_\alpha\in {\sf D}_\alpha$ which should be
always located inside the corresponding domains.}
\label{fi:multid}
\end{figure}

Let now ${\sf D}_{\alpha}$, $\alpha =0, 1, \ldots , g$, be
a {\em collection} of $g+1$ non-intersecting bounded
connected domains in the complex plane
with smooth boundaries $\d {\sf D}_{\alpha}$.
Set ${\sf D}= \cup _{\alpha =0}^{g} {\sf D}_{\alpha}$,
so that the complement ${\sf D^c} = {\bf C}\setminus {\sf D}$ becomes
a multiply-connected unbounded domain in the complex plane
(see fig.~\ref{fi:multid}).
Let ${\sf b}_{\alpha}$ be the boundary curves.
They are assumed to be
positively oriented as boundaries of ${\sf D^c}$, so that
$\cup _{\alpha =0}^{g} {\sf b}_{\alpha} =\d {\sf D^c}$
while ${\sf b}_{\alpha}=-\d {\sf D}_{\alpha}$ has the
clockwise orientation.

Comparing to the simply-connected case, nothing
is changed in posing the standard Dirichlet problem.
The definition of the Green function
and the formula (\ref{Dirih}) for the solution
of the Dirichlet problem through the Green
function remain to be the same.

A difference is in the nature of
harmonic functions. Any harmonic function
is still the real part of an analytic function
but in the multiply-connected case
these analytic funstions are not necessarily single-valued
(only their real parts have to be single-valued).
In other words, the harmonic
functions may have non-zero ``periods" over non-trivial
cycles\footnote{Here and below by ``periods"
of a harmonic function $f$ we mean the integrals
$\oint \d_n f \, dl$ over non-trivial cycles.}. In our case,
the non-trivial cycles are the boundary curves
${\sf b}_{\alpha}$.
In general, the Green function has non-zero ``periods"
over all boundary contours.
Hence it is natural to introduce
new objects, specific to the
multiply-connected case, which are defined
as ``periods"
of the Green function.

First,
the {\it harmonic measure}
$\omega_{\alpha}(z)$
of the boundary component ${\sf b}_{\alpha}$
is the harmonic function in ${\sf D^c}$
such that it is equal to unity on ${\sf b}_{\alpha}$
and vanishes on the other boundary curves.
Thus the harmonic measure is the solution
to the particular Dirichlet problem. From the general
formula (\ref{Dirih}) we conclude that
\be
\label{periodG}
\omega_{\alpha}(z)=
-\, \frac{1}{2\pi}\oint_{{\sf b}_{\alpha}}
\d_n G(z, \zeta )|d\zeta |,
\ \ \ \ \ \ \alpha=1,\ldots,g
\ee
so the harmonic measure is the period of the
Green function w.r.t. one of its arguments. From the maximum
principle for harmonic functions it
follows that $0< \omega_{\alpha}(z)<1$ in internal points.
Obviously, $\sum_{\alpha =0}^{g}\omega_{\alpha}(z)=1$.
In what follows we consider the
linear independent functions
$\omega_{\alpha} (z)$ with $\alpha =1, \ldots , g$.

Further, taking ``periods" in the remaining variable,
we define
\beq\label{periodG2}
\Omega_{\alpha \beta} =- \frac{1}{2\pi}
\oint_{{\sf b}_{\beta}}\d_n \omega_{\alpha} (\zeta )|d\zeta |,
\ \ \ \ \ \ \alpha , \beta =1,\ldots,g
\eeq
The matrix $\Omega_{\alpha \beta}$ is known to be symmetric,
non-degenerate and positively-definite. It will be clear below that
the matrix $T_{\alpha \beta}=i\pi \Omega_{\alpha \beta}$
can be identified with the matrix of periods of holomorphic differentials
on the Schottky double of the domain ${\sf D^c}$
(see formula (\ref{Talbe})). For brevity, we
refer to both $T_{\alpha \beta}$ and $\Omega_{\alpha \beta}$ as
{\em period matrices}.

For the harmonic measure and the period matrix
there are variational formulas similar to the
Hadamard formula (\ref{Hadam}). They can be derived
either by a direct variation of (\ref{periodG}) and
(\ref{periodG2}) using the Hadamard formula or
(much easier) by a ``pictorial'' argument like in
fig.~\ref{fi:hadamar}. The formulas are:
\beq\label{varomega}
\delta \omega_{\alpha}(z)=
\frac{1}{2\pi}\oint_{\d {\sf D}}
\p_n G(z, \xi)\, \p_n \omega_{\alpha}(\xi)\,
\delta n(\xi) \, |d\xi |
\eeq
\beq\label{varperiod}
\delta \Omega_{\alpha \beta}=
\frac{1}{2\pi }\oint_{\d {\sf D}}
\p_n \omega_{\alpha}(\xi)\, \p_n \omega_{\beta}(\xi)
\, \delta n(\xi) \, |d\xi |
\eeq

\paragraph{The Schottky double.}

It is customary to associate with a
planar multiply-connected domain its
{\it Schottky double} (see, e.g.,
\cite{SS}, Ch. 2.2), a compact Riemann surface
without boundary
endowed with antiholomorpic involution, the boundary
of the initial domain being the set of the fixed points
of the involution.
The Schottky double of the
multiply-connected domain ${\sf D^c}$ can be
thought of as two copies
of ${\sf D^c}$ (``upper" and ``lower" sheets
of the double) glued along
the boundaries
$\cup_{\alpha=0}^g {\sf b}_\alpha = \d {\sf D^c}$, with
points at infinity added
($\infty$ and $\bar \infty$).
In this set-up the holomorphic coordinate on
the upper sheet is $z$ inherited from ${\sf D^c}$,
while the holomorphic
coordinate\footnote{More precisely,
the proper
coordinates should be $1/z$ (and $1/\bar z$), which have
first order zeros instead of poles at $z=\infty$ (and
$\bar z=\bar\infty$).}
on the other sheet is $\bar z$.
The Schottky double of
${\sf D^c}$ with two infinities
added is a compact Riemann surface $\Sigma$ of genus
$g = \#\{ {\sf D}_\alpha\}-1$.
A meromorphic function on the double is a pair
of meromorphic functions $f,\tilde f$ on ${\sf D^c}$
such that $f(z)=\tilde f (\bar z)$ on the boundary.
Similarly, a meromorphic differential on the double
is a pair of meromorphic differentials
$f(z)dz$ and $\tilde f (\bar z)d\bar z$ such that
$f(z)dz =\tilde f(\bar z)d\bar z$ along the boundary
curves.

\begin{figure}[tb]
\epsfysize=7.5cm
\centerline{\epsfbox{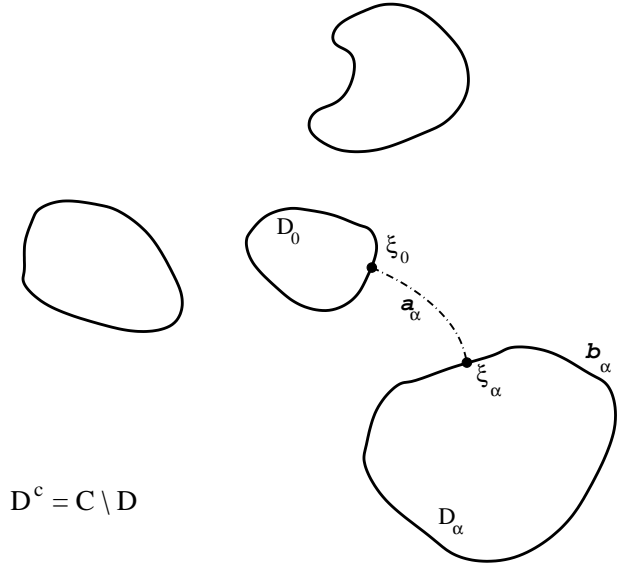}}
\caption{\sl The domain ${\sf D^c}$ with
the ${\sf a}_\alpha$-cycle, going one way along the ``upper
sheet" and back along the ``lower sheet" of the
Schottky double of ${\sf D^c}$.
For such choice one clearly gets the intersection form
${\sf a}_\alpha\circ {\sf b}_\beta = \delta_{\alpha\beta}$ for
$\alpha,\beta=1,\dots,g$.}
\label{fi:Dcut}
\end{figure}

On the double, one may choose a canonical basis
of cycles. We fix the ${\sf b}$-cycles
to be just the boundaries of the
holes ${\sf b}_{\alpha}$ for $\alpha =1, \ldots , g$.
Note that regarded as the oriented boundaries
of ${\sf D^c}$ (not ${\sf D}$) they have the
{\it clockwise} orientation.
The ${\sf a}_{\alpha}$-cycle connects the $\alpha$-th hole
with the 0-th one. To be more precise,
fix points $\xi_{\alpha}$ on the boundaries, then
the ${\sf a}_{\alpha}$ cycle starts from $\xi_{0}$,
goes to $\xi_{\alpha}$ on the ``upper'' (holomorphic)
sheet of the double and goes back
the same way on the ``lower'' sheet, where
the holomorphic coordinate is
$\bar z$, see fig.~\ref{fi:Dcut}.

Being harmonic, $\omega_{\alpha}$ can be represented
as the real part of a holomorphic function:
$$
\omega_{\alpha}(z)=W_{\alpha}(z)+\overline{W_{\alpha}(z)}
$$
where $W_{\alpha}(z)$ are holomorphic multivalued
functions in ${\sf D^c}$. The differentials
$dW_{\alpha}$
are holomorphic in ${\sf D^c}$ and purely imaginary
on all boundary contours. So they
can be extended holomorphically to the lower sheet
as $-d\overline{W_{\alpha}(z)}$.
In fact this is the canonically normalized basis
of holomorphic differentials on the double:
according to the definitions,
$$
\oint_{{\sf a}_{\alpha}}\! dW_{\beta} =
\int_{\xi_0}^{\xi_{\alpha}}
dW_{\beta}(z)+
\int_{\xi_{\alpha}}^{\xi_0}
\left ( - d\overline{W_{\beta}(z)}\right )=
2 {\rm Re} \int_{\xi_0}^{\xi_{\alpha}}
dW_{\beta}(z)=
\omega_{\beta}(\xi_{\alpha})\!-\!
\omega_{\beta}(\xi_0) =\delta_{\alpha \beta}
$$
Then the matrix of ${\sf b}$-periods of these differentials
reads
\be\label{Talbe}
T_{\alpha \beta}=\oint_{{\sf b}_{\alpha}}dW_{\beta} =
- \frac{i}{2}\oint_{{\sf b}_{\alpha}}\p_n \omega_{\beta} dl
=i\pi \Omega_{\alpha \beta}
\ee
i.e. the period matrix $T_{\alpha \beta}$
of the Schottky double $\Sigma$ is a purely
imaginary non-degenerate matrix with positively
definite imaginary part $\pi\Omega_{\alpha \beta}$ (\ref{periodG2}).

\paragraph{Harmonic moments of multiply-connected domains.}

One may still use harmonic moments to characterize
the shape of a multiply-connected domain.
However, the set of harmonic
functions should be extended by adding
functions
with poles in any hole (not only in ${\sf D}_0$
as before) and functions
whose holomorphic parts are not
single-valued. To specify them,
let us mark
points $z_{\alpha}\in {\sf D}_{\alpha}$,
one in each hole (see fig.~\ref{fi:multid}).
Without loss of generality, it is convenient to put
$z_0 =0$. Then one may consider single-valued
analytic functions in ${\sf D^c}$
of the form
$(z-z_{\alpha})^{-k}$ and harmonic functions
$\log \left |1-{z_{\alpha}\over z}\right |^2$ with
multi-valued analytic part.

The arguments almost identical to the ones used in the simply-connected
case show that the parameters
$t_0, M_{n, \alpha}, \phi_{\alpha}$, where as in (\ref{t0})
$t_0 =\mbox{Area}({\sf D})/\pi$,
\be\label{n1}
M_{n,\, \alpha}=
- {1\over \pi}\int_{{\sf D^c}} (z-z_{\alpha})^{-n}d^2 z,
\;\;\;\;\; \alpha =0,1, \ldots , g, \;\; n\geq 1
\ee
together with their complex conjugate, and
\be\label{n0}
\phi_{\alpha}=-{1\over \pi }\int_{{\sf D^c}}
\log \left | 1-\frac{z_{\alpha}}{z}\right |^2
d^2z ,
\;\;\;\;\;
\alpha =1, \ldots , g
\ee
uniquely define ${\sf D^c}$, i.e. any deformation
preserving these parameters is trivial. Note that
the extra moments $\phi_{\alpha}$ are essentially the values
of the potential (\ref{dd1}) at the points $z_{\alpha}$
\be\label{moments1}
\phi_{\alpha}=
\Phi (0) - \Phi (z_{\alpha})-|z_{\alpha}|^2
\ee
A crucial step for what follows is the change of
variables from $M_{n,\alpha}$ to the variables $\tau_k$ which
are finite linear combinations of the $M_{n,\alpha}$'s. They
can be directly defined as moments with respect to new basis of functions:
\be\label{t1}
\tau_0=t_0,\ \ \
{\tau}_{k}={1\over 2\pi i }
\oint_{\p {\sf D}} A_{k}(z)\bar zdz =
-{1\over\pi}\int_{{\sf D}^c}d^2z A_k(z),\ \ \ k> 0
\ee
The functions $A_k(z)$ are analogous to the
Laurent-Fourier type basis on Riemann
surfaces introduced in \cite{kn}.
They are explicitly defined by the following formulas
(here the
indices $\alpha$ and $\beta$ are understood modulo $g+1$):
\be\label{a}
A_{m(g+1)+\alpha}=
R^{-m}(z)\prod_{\beta=0}^{\alpha-1} (z-z_{\beta})^{-1}
\\
R(z)=\prod_{\beta=0}^g (z-z_{\beta})
\ee
In a neighbourhood of infinity $A_k(z)=z^{-k}+O(z^{-k-1})$.
Any analytic function in ${\sf D^c}$ vanishing at infinity
can be represented as a linear combination
of $A_k$ which is {\it convergent} in domains such that
$|R(z)| > {\rm const}$. In the case of one hole ($g=0$) the formulas
(\ref{a})
give the basis used in the previous section: $A_{k}=z^{-k}$.
Note that $A_0 =1$, $A_1 =1/z$, therefore $\tau_0 = t_0$ and
$\tau_1 = M_{1,0}= t_1$.

\paragraph{Local coordinates in the space of
multiply-connected domains.}

Now we are going to prove that the parameters
$\tau_k$, $\phi_{\alpha}$ can be treated as local coordinates
in the space of multiply-connected domains.
(Here we use the same restricted notion of the local coordinates, as
in the simply connected case (see Remark in the Section 2)).

It is instructive and simpler first to prove this
for another choice of parameters.
Instead of $\phi_{\alpha}$ one may use
the areas of the holes
\be\label{dual4}
s_{\alpha}
=\frac{\mbox{Area}({\sf D}_{\alpha})}{\pi}
=\frac{1}{\pi}\int_{{\sf D}_{\alpha}}d^2z =
\frac{1}{2\pi i}\oint_{\p{\sf D}_{\alpha}}
\bar z dz\,,
\;\;\;\;\; \alpha =1, \ldots , g
\ee
In order to prove that any deformation that preserves $\tau_k$ and
$s_{\alpha}$ is trivial,  we introduce
the basis of differentials $dB_k$ which
satisfy the defining ``orthonormality" relations
\be\label{n3}
\frac{1}{2\pi i}
\oint_{\p {\sf D}}A_kdB_{k'}=\delta_{k,k'}
\ee
for all integer $k,k'$. It is easy to see that explicitly they
are given by:
\be\label{o}
dB_{m(g+1)+\alpha}={dz R^{m}(z)\over z-z_g}
\prod_{\beta=0}^{\alpha-1} (z-z_{\beta-1})
\ee
where we identify $z_{-1}\equiv z_g$.
The existence of a well-defined ``dual'' basis of differentials
obeying the orthonormality relation is the key feature
of the basis functions $A_{k}$, which makes $\tau_k$ good local
coordinates comparing to the $M_{n, \alpha}$. For the
functions $(z-z_{\alpha})^{-n}$
one cannot define the dual basis.

The summation formulas
\be
{dz d\zeta \over \zeta -z}=
\sum_{n=1}^{\infty} d \zeta A_{n}(\zeta )
dB_{n}(z), \ \ \
|R(z)| < |R(\zeta)|
\label{n4}\\
{dz d\zeta\over \zeta -z}=
-\sum_{n=0}^{\infty} d\zeta A_{-n}(\zeta )dB_{-n}(z),\ \ \
|R(z)|>|R(\zeta)|
\ee
which can be checked directly, allow us to repeat arguments
of sect.~\ref{ss:simply}.
Indeed, the Cauchy integral (\ref{C}),
\beq\label{CC}
C(z)dz=
{dz\over 2\pi i}\oint_{\p {\sf D}} {\bar \zeta d\zeta
\over \zeta -z}
\eeq
where the integration now goes along all boundary components,
defines in each of the holes ${\sf D_{\alpha}}$ analytic differentials
$C^{\alpha}(z)dz$ (analogs of $C^{+}(z)dz$ in the simply-connected case).
In the complementary domain ${\sf D^c}$ the
Cauchy integral still defines the differential
$C^{-}(z)dz$ holomorphic everywhere in
${\sf D^c}$ except for infinity where it has a simple pole.
The difference of the boundary values of the Cauchy integral
is equal to $\bar z$:
$$
C^{\alpha}(z)-C^{-} (z)=\bar z\,,
\;\;\;\;\;\; z\in \p {\sf D}_{\alpha}
$$                 From equation (\ref{C1}), which can be
written separately for each contour, it follows
that
\begin{itemize}
\item
{\it The difference of the boundary values}
$$\left(\p_t C^{\alpha} (\zeta)-\p_t C^{-}(\zeta)\right)d\zeta$$
{\it of the derivative of the Cauchy integral (\ref{CC})
is, for all $\alpha$,
a purely imaginary differential on the boundary ${\sf b}_{\alpha}$}.
\end{itemize}

\medskip
\noindent
The expansion (\ref{n4}) of the Cauchy kernel implies that
\begin{itemize}
\item
{\it If a $t$-deformation preserves all the
moments $\tau_k$, $k\geq 0$, then
$\p_t \bar\zeta d\zeta-\p_t\zeta d\bar\zeta$
extends to a holomorphic
differential in ${\sf D^c}$}.
\end{itemize}
Indeed, since $|R(z)|$ is small for $z$ close enough
to {\it any} of the points $z_{\alpha}$, one can
expand $\p_t C^{\alpha}(z)$ for any $\alpha$ as
\be\label{n5}
\p_t C^{\alpha}(z)dz={1\over 2\pi i}\sum_{k=1}^{\infty} dB_{k}
\p_t \left (\oint_{\p {\sf D}}
A_{k}(\zeta)\bar \zeta d\zeta \right )=
\sum_{k=1}^{\infty}
\p_t \tau_{k}\ dB_{k}(z)
\ee
and conclude that it is identically zero provided
$\p_t \tau_k =0$. Hence $- \p_t C^{-}(z)dz$ is the desired extension
of $\p_t \bar\zeta d\zeta-\p_t\zeta d\bar\zeta$. It has no
pole at infinity due to the equation $\p_t \tau_0=0$.

Using the Schwarz symmetry principle we obtain that $\p_t C^{-}(z)dz$ extends to
a holomorphic differential on the Schottky double. If the variables $s_{\alpha}$
are also preserved under the $t$-deformation,
then this holomorphic differential has
zero periods along all the cycles ${\sf b}_{\alpha}$. Therefore,
it is identically zero. This completes the proof of the statement that
any deformation of the domain preserving
$\tau_k$ and $s_{\alpha}$ is trivial.

In this proof the variables $s_{\alpha}$ were used
only at the last moment
in order to show that the extension of $\p_t C^{-}(z)dz$
as a holomorphic differential
on the Schottky double is trivial.
The variables $\phi_{\alpha}$ can be used
in a similar way. Namely, let us show that if they are preserved under
$t$-deformation then ${\sf a}_{\alpha}$-periods of the extension of
$\p_t C^{-}(z)dz$ are trivial,
and therefore this extension is identically zero.
Indeed, the variable $\phi_{\alpha}$
(\ref{n0}) can be represented in the form
\be\label{n20}
\phi_{\alpha}=-{2\over\pi}{\rm Re}\int_{0}^{z_{\alpha}}
dz\int_{{\sf D^c}}
\frac{d^2 \zeta}{z-\zeta}
\ee
The differential
$\frac{dz}{\pi}\int_{{\sf D^c}}
\frac{d^2 \zeta}{z-\zeta}$ is equal to
$C^{\alpha}(z)dz$ for $z\in {\sf D}_{\alpha}$
and $(\bar z + C^{-}(z))dz$ for $z\in {\sf D^c}$.
Let $\xi_0$, $\xi_{\alpha}$ be the points where the
integration path from $0$ to $z_{\alpha}$ intersects
the boundary contours ${\sf b}_0$, ${\sf b}_{\alpha}$. Then
\be\label{n201}
\phi_{\alpha}=-2\, {\rm Re} \left(
\int_{0}^{\xi_0}C^{0} (z)dz +
\int_{\xi_{\alpha}}^{z_{\alpha}}C^{\alpha} (z)dz +
\int_{\xi_0}^{\xi_{\alpha}}(\bar z + C^{-} (z))dz \right)
\ee
It is shown above that if
a $t$-deformation preserves the variables
$\tau_k$ then all $\p_t C^{\alpha}(z)dz=0$.
Thus vanishing of the $t$-derivative $\p_t \phi_{\alpha} =0$ implies
\be\label{n21}
0=- \p_t \phi_{\alpha} =
2\, {\rm Re}\int_{\xi_0}^{\xi_{\alpha}} \p_tC^{-}(z)dz
\ee
The r.h.s of this equation is just the
${\sf a}_{\alpha}$-period of the holomorphic extension
of the differential $\p_tC^{-}(z)dz$.

Let us construct the deformations
$\p_{x_k}^{\phi}$ and $\p_{y_k}^{\phi}$
of the boundary that change the real
or imaginary parts of the variable
$\tau_k=x_k+iy_k$,
$k\geq 1$, keeping all the other
moments and the variables $\phi_{\alpha}$ fixed.
It is convenient to set $\p_{\tau_k}^{\phi}=\frac{1}{2}
(\p_{x_k}^{\phi}-i \p_{y_k}^{\phi})$.
The argument is similar to the proof
of the fact that any deformation that preserves all
the variables is trivial.

\medskip
\begin{itemize}
\item{\it Suppose that the deformations
$\p_{x_k}^{\phi}$ and $\p_{y_k}^{\phi}$ exist.
Then the differential
$\p_{\tau_k}^{\phi} \bar\zeta d\zeta-\p_{\tau_k}^{\phi}\zeta d\bar\zeta$
extends from $\p{\sf D^c}$ to the Schottky double $\Sigma$. Its extension is
a meromorphic differential
$d\Omega_k^{\phi}$ with the only pole at the infinity
point $\infty$ on the upper sheet. In a neighborhood of $\infty $
it has the form}
\be\label{n25}
d\Omega_k^{\phi}(z)=dB_{k}(z)+O(z^{-2})dz
\ee
{\it The ${\sf a}$-periods of $d\Omega_k^{\phi}$ are equal to}
\beq\label{n26}
\oint_{{\sf a}_{\alpha}}d\Omega_k^{\phi}=\int_0^{z_{\alpha}} dB_k
\eeq
\end{itemize}
First of all, it is clear that
the meromorphic differential
$d\Omega_k^{\phi}$ on $\Sigma$ is uniquely defined
by its asymptotics at $\infty$
and by the normalization (\ref{n26}) of its ${\sf a}$-periods.
To deduce these properties,
we notice, using (\ref{n5}), that
$\p_{x_k}C^{\alpha}(z)dz=dB_{k}(z)$.
Therefore, the differential
$\p_{x_k}^{\phi} \bar\zeta d\zeta-\p_{x_k}^{\phi}\zeta d\bar\zeta$
extends to ${\sf D^c}$ as
\be\label{n28}
d\Omega_{x_k}^{\phi}=-\p_{x_k} C^{-}(z)dz+dB_{k}
\ee
Using the Schwarz symmetry principle we conclude
that it extends to
the Schottky double as a meromorphic differential.
Around the two infinities
it has the form
$d\Omega_{x_k}^{\phi}\ \stackreb{z\to\infty}{=}\ dB_k+O(z^{-2})dz$ and
$d\Omega_{x_k}^{\phi}\ \stackreb{\bar z\to\bar\infty}{=}\
-d\bar B_k+O(\bar z^{-2})d\bar z$.
In the same way one gets that the differential
$\p_{y_n}^{\phi} \bar\zeta d\zeta-\p_{y_n}^{\phi}\zeta d\bar\zeta$
extends to the double as a meromorphic differential $d\Omega_{y_k}^{\phi}$,
which at the two infinities has the form
$d\Omega_{y_k}^{\phi}\ \stackreb{z\to\infty}{=}\ idB_k+O(z^{-2})dz$ and
$d\Omega_{y_k}^{\phi}\ \stackreb{\bar z\to\bar\infty}{=}\
id\bar B_k+O(\bar z^{-2})d\bar z$ respectively.
Since
$$
2d\Omega_{k}^{\phi}=d\Omega_{x_k}^{\phi}-id\Omega_{y_k}^{\phi}
$$
the first statement is proven. From
$\p_{x_k}^{\phi} \phi_{\alpha}=0$ and (\ref{n20}), (\ref{n28})
it follows that
$$
0={\rm Re}\left(\int_0^{\xi_0}dB_{k}+
\int_{\xi_0}^{\xi_{\alpha}}
\left(-d\Omega_{x_k}^{\phi}+dB_{k}\right)+
\int_{\xi_{\alpha}}^{z_{\alpha}}dB_{k}\right)
$$
Hence
$$
\oint_{{\sf a}_{\alpha}}d\Omega_{x_k}^{\phi}=2{\rm Re} \int_0^{z_{\alpha}}dB_k
$$
In the same way one gets
$$
\oint_{{\sf a}_{\alpha}}d\Omega_{y_k}^{\phi}=
-2{\rm Im} \int_0^{z_{\alpha}}\,dB_k
$$
The last two equations are equivalent to (\ref{n26}).

Normal displacement of the boundary that accomplishes
the deformations can be explicitly
found using the following elementary proposition.
\begin{itemize}
\item
{\it Let ${\sf D}(t)$ be a deformation with real parameter $t$
such that the differential}
$$
d\Omega = \p_t \bar \zeta d\zeta -\p_t \zeta d\bar \zeta
$$
{\it extends to a meromorphic differential $d\Omega$
globally defined on the Schottky double $\Sigma$.
Then the corresponding normal displacement of the
boundary is proportional to normal derivative of
${\rm Re} \int^{z} d\Omega $ at the boundary point $\xi$:}
\beq\label{Omeganormal}
\delta n(\xi )=
\frac{1}{2}
\delta t\, \p_n \left ( {\rm Re} \int^{\xi}
d\Omega \right )
\eeq
{\it Conversely, if $\delta n(\xi )=\frac{1}{2}\delta t \,
\p_n H(\xi )$, where $H$ is a real-valued function
such that $dH=0$ along the boundary contours and
$\p_z H$ is meromorphic in ${\sf D^c}$ then
the differential
$\p_t \bar \zeta d\zeta -\p_t \zeta d\bar \zeta$
is meromorphically extendable to the Schottky double
as $2\p_z H dz$ on the upper sheet and
$-2\p_{\bar z}H d\bar z$ on the lower sheet.}
\end{itemize}
In our case normal displacements
of the boundary that change
$x_k$ or $y_k$ keeping all the other moments
and the variables $\phi_{\alpha}$ fixed are thus
given by
\beq\label{normaldisp}
\delta n(\xi )=
\frac{1}{2}
\delta x_k\, \p_n \left ( {\rm Re} \int^{\xi}
d\Omega_{x_k}^{\phi} \right ),
\;\;\;\;
\delta n(\xi )=
\frac{1}{2}
\delta y_k\, \p_n \left ( {\rm Re} \int^{\xi}
d\Omega_{y_k}^{\phi} \right )
\eeq
respectively. Note that
since the differentials $d\Omega_{x_k}(z)$, $d\Omega_{y_k}(z)$
(but not $d\Omega_{k}(z)$!) are purely imaginary on the
boundaries,
$d\, {\rm Re} \, \Omega_{x_k}(z)=d\, {\rm Re} \,
\Omega_{y_k}(z)=0$
along each component of the boundary. With
formulas (\ref{normaldisp}) at hand,
one can directly verify that these deformations
indeed change $x_k$ or $y_k$ only
and keep fixed all other moments.
We leave this to the reader.

In terms of the differential $d\Omega_k^{\phi}$
formulas (\ref{normaldisp}) acquire the form
\beq\label{normaldisp1}
\delta n(\xi )=
\delta x_k\, \p_n \left ( {\rm Re} \int^{\xi}
d\Omega_{k}^{\phi} \right ),
\;\;\;\;
\delta n(\xi )=
- \delta y_k\, \p_n \left ( {\rm Im} \int^{\xi}
d\Omega_{k}^{\phi} \right )
\eeq
(cf. (\ref{Hka}) for the
simply-connected case). Indeed, taking the real part of
$2\Omega_k (\xi )=\Omega_{x_k}(\xi)-i \Omega_{y_k}(\xi )$,
we get $2 \, {\rm Re} \, \Omega_k (\xi )=\, {\rm Re} \,
\Omega_{x_k}(\xi)
+\, {\rm Im} \, \Omega_{y_k}(\xi )$. But the normal
derivative of ${\rm Im} \, \Omega_{y_k}(\xi )$ vanishes since,
by virtue of the Cauchy-Riemann identities, it is equal
to the tangential derivative of the conjugate harmonic
function ${\rm Re} \, \Omega_{y_k} (\xi )$. This proves
the first formula in (\ref{normaldisp1}). The second one
is proven in a similar way by taking imaginary part
of $2\Omega_k (\xi )$.

The construction of the vector fields $\p_{\tau_0}^{\phi}$
(which changes $\tau_0$ only) and
$\p_{\alpha}^{\phi}$ (which changes $\phi_{\alpha}$ only)
is quite similar and even simpler since the derivative
(\ref{n5}) vanishes. So, we
present the results without going into details.

\medskip
\begin{itemize}
\item{\it The deformation $\p_{\tau_0}^{\phi}$ corresponds
to the normal displacement}
$$
\delta n (\xi ) =-\frac{1}{2}\delta \tau_0 \, \p_n
G(\infty , \xi )
$$
{\it The differential
$-(\p_{\tau_0}^{\phi} \bar\zeta d\zeta-
\p_{\tau_0}^{\phi}\zeta d\bar\zeta )$
extends from $\p{\sf D^c}$ to the Schottky double
$\Sigma$. Its extension is
a meromorphic third-kind Abelian differential
$d\Omega_0$ which has simple poles
at the infinities on the two sheets of the Schottky double
(with residues $\pm 1$) and vanishing ${\sf a}$-periods.}
\item{\it The deformation $\p_{\alpha}^{\phi}$ corresponds
to the normal displacement}
$$
\delta n (\xi ) =\frac{1}{4}\delta \phi_{\alpha} \, \p_n
\omega_{\alpha}(\xi )
$$
{\it where
$\omega_{\alpha}$ is the harmonic measure of the
boundary component ${\sf b}_{\alpha}$
(see (\ref{periodG})).
The differential
$\p_{\alpha}^{\phi} \bar\zeta d\zeta-
\p_{\alpha}^{\phi}\zeta d\bar\zeta$ holomorphically
extends from $\p{\sf D^c}$ to the Schottky double $\Sigma$.
Its extension is the canonically normalized
holomorphic differential
$dW_{\alpha}=\p_z \omega_{\alpha}(z)dz$ on the upper sheet
(and $dW_{\alpha}=-\p_{\bar z} \omega_{\alpha}(z)d\bar z$
on the lower sheet).}
\end{itemize}

So we see that
$\p_{x_k}^{\phi}$, $\p_{y_k}^{\phi}$,
$\p_{\tau_0}^{\phi}$ and $\p_{\alpha}^{\phi}$
are well-defined
vector fields on the space of multiply-connected domains.
This fact allows us to treat $\phi_{\alpha}$, $\tau_k$
as local coordinates on this space.
At this stage it becomes clear why we prefer to use
the moments $\tau_k$ rather than $M_{k,\alpha}$.
Although the latter are finite linear combinations
of the former, they can not be treated as local coordinates
because the vector fields $\p/\p M_{k,\alpha}$, being
in general {\it infinite} linear combinations of
the $\p_{\tau_k}^{\phi}$, are not well-defined.

\paragraph{$\Pi$-variables.}
Up to now the roles of the variables $s_{\alpha}$ and $\phi_{\alpha}$
have been in some sense dual to each other.
It is necessary to emphasize that this duality does not go beyond the
framework of our proof of the statement that the first or the second sets
together with the variables $\tau_k$ are local coordinates in the
space of multiply-connected domains. For analytic boundary curves
one can define the Schwarz function, which is a
unique function
analytic in some strip-like neighborhoods of all boundaries
such that
\beq\label{Schwarz}
S(z)=\bar z \;\;\;\;\;
\mbox{on the boundary curves}
\eeq
Then the variables $s_{\alpha}$
are ${\sf b}$-periods of the differential
$S(z)dz$. At the same time, the variables $\phi_{\alpha}$
in general can not be identified with periods
of this differential (or its extension)
over any cycles on the Schottky double.
Now we are going to introduce
new variables, $\Pi_{\alpha}$, which can be called
{\it virtual} ${\sf a}$-periods
of the differential $S(z)dz$ on the Schottky
double, since in all the cases
when the Schwarz function has a meromorphic extension
to the double they indeed coincide
with the ${\sf a}$-periods of the corresponding differential
(see below in this section).

Let us consider the differential
\be\label{n32}
d\Omega_k=d\Omega_{k}^{\phi}-\sum_{\alpha}B_k
(z_{\alpha})dW_{\alpha}
\;\;\;\;\; (k\geq 1)
\ee
where
\be\label{n32a}
B_k(z)=\int_0^zdB_{k}
\ee
is a polynomial of degree $k$.
It is a meromorphic differential on $\Sigma$ with the only pole at
$\infty$ on the upper sheet, where it has the form
\be\label{n25a}
d\Omega_k(z)=dB_{k}(z)+O(z^{-2})dz
\ee           From (\ref{n26}) it is clear that
the differential $d\Omega_k$ has vanishing ${\sf a}$-periods
\be\label{n26a}
\oint_{{\sf a}_{\alpha}}d\Omega_k=0
\ee
i.e. it is a canonically
normalized meromorphic differential.
The normal displacements of the boundary given by real and imaginary parts of the
normal derivative $\p_n \Omega_k$ define a complex tangent vector field
\be\label{n33}
\p_{\tau_k}^{\Pi}=\p_{\tau_k}^{\phi}-
\sum_{\alpha}B_k(z_{\alpha})\p^{\phi}_{\alpha}
\ee
to the space of multiply-connected domains.
These vector fields keep fixed the formal variable
\be\label{n34}
\Pi_{\alpha}=\phi_{\alpha}+2{\rm Re}\sum_{k}B_k(z_{\alpha})\tau_k
\ee
In general situation
this variable is only a formal one
because the sum generally does not converge.
Thus, we call $\Pi_{\alpha}$ the
{\it virtual} ${\sf a}$-period
of the Schwarz differential $S(z)dz$, since
in the case when the Schwarz function has
a meromorphic extension to the double $\Sigma$, the sum does converge
and the corresponding quantity does coincide with the
${\sf a}_{\alpha}$-period of
the extension of the Schwarz differential.

\paragraph{Elementary deformations and the operator $\nabla(z)$.}
Like in sect. 2, we introduce the elementary deformations
\be
\label{dispmu}
\delta_a
\;\;\;\mbox{with}\;\;\;\;
\delta n (\xi )=-\, \frac{\epsilon}{2}
\p_n G(a, \xi )\,,
\;\;\;\; a \, \in \, {\sf D^c}
\\
\delta^{(\alpha )}
\;\;\mbox{with}\;\;\;\;
\delta n (\xi )=-\, \frac{\epsilon}{2}
\p_n \omega_{\alpha}(\xi)\,,
\;\;\;\; \alpha =1, \ldots , g
\ee
where $\omega_{\alpha}(z)$ is the harmonic measure of the boundary
component ${\sf b}_{\alpha}$ (see (\ref{periodG})).
The deformations $\delta^{(\alpha )}$ were considered in
\cite{Gustafsson} in connection with so called quadrature domains
\cite{Gustafsson,Ahar-Shap}.

In complete analogy with sect. 2 one can derive
the following formulas for variations of the local
coordinates under elementary deformations:
\beq\label{elem1}
\delta_a \tau_k =\epsilon A_k(a),
\;\;\;\;
\delta_a \phi_{\alpha}=
\epsilon \log \left |1\! -\! \frac{z_{\alpha}}{a}\right |^2,
\;\;\;\;
\delta^{(\alpha )}\tau_k =0,
\;\;\;\;
\delta^{(\alpha )}\phi_{\beta} =-2\epsilon
\delta_{\alpha \beta}
\eeq
The first two formulas are particular cases of
$$
\delta_a \int_{{\sf D^c}}  h(\zeta )d^2 \zeta
=\frac{\epsilon}{2} \oint_{\p {\sf D^c}} h(\zeta )
\p_n G(a, \zeta )|d\zeta | =-\epsilon \pi h(a)
$$
which is
valid for any harmonic function $h$ in ${\sf D^c}$
(the last equality is just the formula for
solution of the Dirichlet problem).
Similarly,
$$
\delta^{(\alpha )}
\int_{{\sf D^c}}  h(\zeta )d^2 \zeta
=\frac{\epsilon}{2} \oint_{\p {\sf D^c}} h
\p_n \omega_{\alpha} \, |d\zeta |=
\frac{\epsilon}{2} \oint_{{\sf b}_{\alpha}}
\p_n h \, |d\zeta |=-i\epsilon
\oint_{\p {\sf D}_{\alpha}}
\p_{\zeta}h \, d\zeta
$$
(the Green formula was used), and the last two formulas
in (\ref{elem1}) correspond to the particular choices
of $h(z)$.
Variations of the variables $\Pi_{\alpha}$
(in the case when they are well-defined) then read:
\beq\label{elem1aaa}
\delta_z \Pi_{\alpha}=0,
\;\;\;\;
\delta^{(\alpha )}\Pi_{\beta} =-2\epsilon
\delta_{\alpha \beta}
\eeq

Therefore, for any functional $X$
on the space of the multiply-connected domains
the following equations hold:
\be\label{m6}
\delta_z X=\epsilon\nabla(z)X
\ee
\be\label{n40}
\delta^{(\alpha)}X =-2\epsilon \p_{\alpha}^{\phi}X=-2\epsilon
\p_{\alpha}^{\Pi}X
\ee
The differential operator $\nabla(z)$ in the
multiply-connected case is defined
by the formula
\be\label{D2m}
\nabla (z)
=\p_{\tau_0}^{\Pi} +\sum_{k\geq 1}
\left( A_k(z)\p_{\tau_k}^{\Pi} +
\overline{A_k(z)}\p_{\bar \tau_k}^{\Pi}\right )
\ee
The functional $X$ can be regarded as a function
$X=X^\phi(\phi_{\alpha},\tau_k)$
on the space of the local coordinates
$\phi_{\alpha},\tau_k$, or as a function $X=X^\Pi(\Pi_{\alpha},\tau_k)$
on the space of the local coordinates $\Pi_{\alpha},\tau_k$.
We would like to stress once again, that although in the latter case the
variables $\Pi_{\alpha}$ are formal their variations under elementary deformations
and the vector-fields
$\p_{\tau_k}^{\Pi}$, which keep them fixed, are well-defined.

For completeness, let us characterize elementary
deformations $\delta_a$ in terms of meromorphic
differentials on the Schottky double (as we have
already seen, deformations $\delta^{(\alpha )}$
correspond to holomorphic differentials).
\begin{itemize}
\item{\it
Let $\p_{t^{(a)}}$ be the vector field in the space of
multiply-connected domains corresponding to the
elementary deformation $\delta_a$. Then
the differential
$-(\p_{t^{(a)}} \bar\zeta d\zeta-
\p_{t^{(a)}}\zeta d\bar\zeta )$
extends from $\p{\sf D^c}$ to the Schottky double $\Sigma$.
Its extension is
a meromorphic third-kind Abelian differential
$d\Omega^{(a, \bar a)}$ which has simple poles
at the points $a$ and $\bar a$ on the two sheets of the Schottky double
(with residues $\pm 1$) and vanishing ${\sf a}$-periods.}
\end{itemize}
In terms of the Green function we have:
$$
d\Omega^{(a, \bar a)}=\left \{
\begin{array}{c}
2 \p_z G(a,z)dz \;\;\;\;\mbox{on the upper sheet}
\\ \\
-2 \p_{\bar z} G(a,z)dz \;\;\;\;\mbox{on the lower sheet}
\end{array}
\right.
$$
(cf. (\ref{Omeganormal}) and (\ref{dispmu})).
Note that the differential $d\Omega_0$ introduced before
coincides with $d\Omega^{(\infty , \bar \infty )}$.

Let $K(z, \zeta)d\zeta$
be a unique meromorphic Abelian differential of the third kind
on $\Sigma$ with simple poles at $z$ and $\infty$
on the upper sheet with residues $\pm 1$
normalized to zero ${\sf a}$-periods.
(Note that as a function of the variable $z$
it is multi-valued on $\Sigma$.)
Then
\be\label{m5}
2\p_{\zeta}G(z,\zeta)d\zeta - 2\p_{\zeta}G(\infty,\zeta)d\zeta
=K(z,\zeta)d\zeta +K(\bar z,\zeta)d\zeta
\ee
and the differential $d\Omega_k(\zeta)$ can be represented in the form
\be\label{m4}
d\Omega_k(\zeta)={d\zeta \over 2\pi i}
\oint_{\infty} K(u,\zeta) dB_k(u)
\ee
where the $u$-integration goes along a big circle
around infinity.
Using (\ref{n4}) we obtain that
\be\label{m41}
-\sum_{k\geq 1} A_k(z)d\Omega_k(\zeta)=
{d\zeta \over 2\pi i}\oint_{\infty}
{K(u,\zeta)du\over u-z}=K(z,\zeta)d\zeta
\ee
Therefore, the following expansion of the derivative
of the Green function holds:
\beq\label{expder}
2\p_{\zeta}G(z, \zeta )d\zeta =d\Omega_0 (\zeta)-
\sum_{k\geq 1}\left (
A_k (z)d\Omega_k (\zeta)+\overline{A_k(z)}
d\bar \Omega_k (\zeta )\right )
\eeq
Here $d\bar \Omega_k$ is a unique meromorphic
differential on $\Sigma$ with the only pole
at infinity on the lower sheet with the principal
part $-d\overline{B_k(z)}$ and
vanishing ${\sf a}$-periods.
This formula generalizes
eq.\,(3.8) from \cite{MWZ} to the
multiply-connected case.

\paragraph{The $F$-function.}
Applying the variational formulas (\ref{Hadam}),
(\ref{varomega}), (\ref{varperiod}), we can
find variations of the Green function, harmonic measure
and period matrix under the elementary deformations.
In this way we obtain a number of important
relations which connect elementary deformations
of these objects:
\beq\label{elem3}
\begin{array}{l}
\delta_a G(b,c) =\delta_b G(c,a) =\delta_c G(a,b)
\\ \\
\delta_a \omega_{\alpha}(b)=\delta^{(\alpha )}G(a,b)
=\delta_b \omega_{\alpha}(a)
\\ \\
\delta^{(\alpha)}\omega_{\beta}(z)=
\delta^{(\beta)}\omega_{\alpha}(z)
\\ \\
\delta_z \Omega_{\alpha \beta}=
\delta^{(\alpha)}\omega_{\beta}(z)
\\ \\
\delta^{(\alpha)}\Omega_{\beta \gamma}=
\delta^{(\beta)}\Omega_{\gamma \alpha}=
\delta^{(\gamma)}\Omega_{\alpha \beta}
\end{array}
\eeq    From (\ref{m6}), (\ref{n40})
it follows that
the formulas (\ref{elem3}) can be rewritten in terms
of the differential operators $\nabla (z)$ and
$\p_{\alpha}:=\p / \p\phi_{\alpha}=\p/\p\Pi_{\alpha}$:
\beq\label{elem4}
\begin{array}{l}
\nabla (a) G(b,c) =\nabla (b) G(c,a) =\nabla (c) G(a,b)
\\ \\
\nabla (a)\omega_{\alpha}(b)=-2\p_{\alpha} G(a,b)
\\ \\
\p_{\alpha}\omega_{\beta}(z)=
\p_{\beta} \omega_{\alpha}(z)
\\ \\
\nabla (z) \Omega_{\alpha \beta}=
-2\p_{\alpha} \omega_{\beta}(z)
\\ \\
\p_{\alpha}\Omega_{\beta \gamma}=
\p_{\beta} \Omega_{\gamma \alpha}=
\p_{\gamma} \Omega_{\alpha \beta}
\end{array}
\eeq
These integrability relations generalize
formulas (\ref{symha}) to the multiply-connected case.
The first line just coincides with (\ref{symha})
while the other ones extend the symmetry of the
derivatives
to the harmonic measure and the period matrix.

Again, (\ref{elem4}) can be regarded
as a set of compatibility
conditions of an infinite hierarchy of
differential equations.
They imply that there exists a function
$F = F(\Pi_{\alpha}, \btau)$ such that
\beq\label{elem5}
G(a,b)=\log \left |a^{-1}-b^{-1}\right |+
\frac{1}{2}\nabla (a)\nabla (b)F
\eeq
\beq\label{elem51}
\omega_{\alpha}(z)=-\, \p_{\alpha}\,\nabla (z)F
\eeq
\beq\label{elem52}
T_{\alpha \beta}=i \pi \Omega_{\alpha \beta}=2\pi i \,
\p_{\alpha}\p_{\beta} F
\eeq
The function $F$ is the (logarithm of the)
tau-function of multiply-connected domains.

\paragraph{Dual moments and integral representation
of the tau-function.}

To obtain the integral representation of the
function $F$,
we proceed exactly in the same manner as in
sect.~\ref{ss:simply} (see also \cite{MWZ} for more details).

Again, set $\tilde \Phi (z)=\nabla (z)F$.
Eqs.\,(\ref{elem5}) and (\ref{elem51}) determine the function
$\tilde \Phi (z)$
for $z\in {\sf D^c}$ via its variations under the elementary
deformations:
\beq\label{dual1}
\begin{array}{l}
\displaystyle{
\delta_a \tilde \Phi (z) =-2\epsilon
\log \left |a^{-1}-z^{-1}\right | +2\epsilon G(a,z)}
\\ \\
\delta^{(\alpha)}\tilde \Phi (z)=
2\epsilon \omega_{\alpha}(z)
\end{array}
\eeq
It is easy to verify that the function
\beq\label{modpot1}
\tilde \Phi (z)=-\frac{2}{\pi}\int_{{\sf D}}
\log |z^{-1}-\zeta^{-1}|d^2 \zeta =
\Phi (z)-\Phi (0)+\tau_0 \log |z|^2
\eeq
satisfies (\ref{dual1}). Indeed, using (\ref{dispmu}), variation of
(\ref{modpot1}) reads
$$
\delta_a \left(-\frac{2}{\pi}\int_{{\sf D}}
\log |z^{-1}-\zeta^{-1}|d^2 \zeta\right) = {\epsilon\over\pi}
\oint_{\p{\sf D^c}} |d\xi|\p_n G(a,\xi) \log |z^{-1}-\xi^{-1}|
$$
$$
={\epsilon\over\pi}
\oint_{\p{\sf D^c}} |d\xi|\p_n G(a,\xi)\left(\log |z^{-1}-\xi^{-1}|
-G(z,\xi)\right) = -2\epsilon
\log \left |a^{-1}-z^{-1}\right | +2\epsilon G(a,z)
$$
where we have used properties of the Dirichlet Green function and the
fact that Dirichlet formula restores harmonic function from its value at the
boundary. Similarly, for $z \in {\sf D^c}$ we obtain:
$$
\delta^{(\alpha)} \left(-\frac{2}{\pi}\int_{{\sf D}}
\log |z^{-1}-\zeta^{-1}|d^2 \zeta\right) = {\epsilon\over\pi}
\oint_{\p{\sf D^c}} |d\xi|\p_n
\omega_\alpha(\xi) \log |z^{-1}-\xi^{-1}|
$$
$$
={\epsilon\over\pi}
\oint_{\p{\sf D^c}} |d\xi|\p_n \omega_\alpha(\xi)
\left(\log |z^{-1}-\xi^{-1}| -G(z,\xi)\right)
$$
$$
={\epsilon\over\pi}
\oint_{{\sf b}_\alpha} |d\xi|\p_n \left(\log |z^{-1}-\xi^{-1}|
-G(z,\xi)\right) =
2\epsilon \omega_\alpha(z)
$$
The same calculation for $z\in {\sf D}$ yields
\beq\label{zind}
\delta^{(\alpha )}\tilde \Phi (z)=
\left \{
\begin{array}{cl}
0 & \;\;\mbox{if $z\in {\sf D}_0$}
\\&\\
2\epsilon \delta_{\alpha \beta}
&\;\;\mbox{if $z\in {\sf D}_{\beta}$}\, ,
\;\;\beta =1, \ldots , g
\end{array}\right.
\eeq

We see that expression in (\ref{modpot1})
coincides with
$\tilde \Phi$ given by (\ref{modpot}),
where ${\sf D}$ is now understood as the union of
all ${\sf D}_{\alpha}$'s.
The coefficients of an expansion of $\tilde \Phi$ at infinity
define the dual moments $\nu_k$:
\be\label{n50}
\nabla (z)F =\tilde \Phi (z)=
-\frac{2}{\pi} \int_{{\sf D}} \log |z^{-1}-\zeta^{-1}| d^2 \zeta
= v_0 + 2\Re\sum_{k>0}\nu_k A_k(z)
\ee
The coefficients in the r.h.s. of (\ref{n50}) are moments of the
union of the interior domains with respect to the
dual basis
\be\label{n51}
\nu_k={1\over \pi}\int_{{\sf D}} B_k(z)d^2z
\ee                 From equation (\ref{n50}) it follows that
\be\label{n52}
\nu_k=\p_{\tau_k}^\Pi F
\ee
The same arguments show that the derivatives
\beq\label{dual2}
s_{\alpha}:=-\, \p_{\alpha} F
\eeq
are just areas of the holes (\ref{dual4}). Indeed,
eqs.\,(\ref{elem51}), (\ref{elem52}) determine these quantities
via their variations:
$\delta_a s_{\alpha}=\epsilon \omega_{\alpha}(a)$,
$\delta^{(\beta)}s_{\alpha}=\epsilon \Omega_{\alpha \beta}$.
A direct check, using (\ref{dispmu}), shows that
\be\label{dual5}
\nabla (z)F= -\frac{2}{\pi} \int_{{\sf D}} \log
\left |\frac{1}{z}-\frac{1}{\zeta}\right |
d^2 \zeta\,,
\;\;\;\;\;\;\;
\p_{\alpha} F=
-\, \frac{\mbox{Area}({\sf D}_{\alpha})}{\pi}
\ee
For example,
$$
\delta^{(\alpha)}F = \frac{1}{2\pi}
\delta^{(\alpha)} \left(\int_{{\sf D}}
\tilde\Phi(z) d^2z\right) = -{\epsilon\over 4\pi}
\oint_{\p{\sf D^c}} |d\xi|\p_n \omega_\alpha(\xi)\tilde\Phi(\xi)
+\frac{1}{2\pi} \int_{{\sf D}} \delta^{(\alpha )}\tilde
\Phi (\zeta ) \, d^2 \zeta
$$
In the last term we use (\ref{zind}) and obtain the result:
$$
\delta^{(\alpha)}F =\, -{\epsilon\over 4\pi}
\oint_{{\sf b}_\alpha} |d\xi|\p_n\tilde\Phi(\xi)
+\frac{\epsilon}{\pi}\int_{{\sf D}_{\alpha}}d^2 \zeta
=-\frac{\epsilon}{4\pi}\int_{{\sf D}_{\alpha}}\Delta
\tilde \Phi \, d^2 \zeta +\epsilon s_{\alpha}=2\epsilon s_{\alpha}
$$
(here $\Delta =4\p_z \p_{\bar z}$ is the Laplace operator).

The integral representation of $F$ is found
in the same way through its variations which
are read from (\ref{dual5}). The result is given
by the same formulas (\ref{F}) and (\ref{F0}) as in the
simply-connected case with the understanding
that ${\sf D}=\cup_{\alpha=0}^g{\sf D}_{\alpha}$ is now
the union of all ${\sf D}_{\alpha}$'s.

\paragraph{Algebraic domains.}

In what follows we restrict our analysis
by the class of {\it algebraic domains}.
In the simply-connected case
dealt with in the previous section
the algebraic domains are simply
images of the exterior of the unit disk under
one-to-one conformal maps given by {\it rational functions}
whose singularities are all in the other ``half''
of the plane, i.e. inside the unit circle.
Note that the boundary of the unit circle is the set of fixed points
of the inversion $w\to 1/\bar w$ which is the antiholomorphic
involution of the $w$-plane compactified by a point at infinity
(the Riemann sphere).

Planar multiply-connected algebraic domains
can be defined as the domains
for which the Schwartz function has a meromorphic
extension to a higher genus Riemann surface (a complex
algebraic curve) with antiholomorphic
involution.
More precisely, let $\Sigma$ be a {\it real} Riemann
surface by which we mean a complex
algebraic curve
of genus $g$ with an antiholomorphic involution such that
the set of fixed points consists of exactly $g+1$
closed contours (such curves are sometimes called $M$-curves).
Then $\Sigma$ can be naturally
divided in two ``halves'' (say upper and lower sheets)
which are interchanged by the involution.
Algebraic domains with $g$ holes in the plane
can be defined as images of the upper half of the real
Riemann surface under bijective conformal maps given
by rational (meromorphic) functions on $\Sigma$
(see, e.g. \cite{EV}).
For the purpose of this paper,
it is convenient to use another, more
direct characterization of algebraic domains.

The domain ${\sf D^c}$ is algebraic if and only if
the Cauchy integrals (\ref{CC})
$$
C^{\alpha}(z)=\frac{1}{2\pi i}
\oint_{\p {\sf D}}\frac{\bar \zeta d\zeta}{\zeta -z}
\;\;\;\mbox{for $z\in {\sf D}_{\alpha}$}
$$
are extendable to a rational (meromorphic) function $J(z)$
on the whole complex plane with a marked point at
infinity (see \cite{EV}). It is important to stress
that this function is required to be the same for all
$\alpha$.
The equality
$$
S(z)=J(z)-C^{-}(z)
$$
valid by definition for $z\in \partial {\sf D^c}$ can be used for
analytic extension of the Schwarz function.
The function $C^{-}(z)$ is analytic in
${\sf D^c}$. Therefore, $J(z)$ and $S(z)$ have the same singular parts at
their poles in ${\sf D^c}$.
One may treat $S(z)$ as a function on the
Schottky double extending it to the lower sheet as
$\bar z$.

It is also convenient to introduce
\be\label{alg2}
V(z)=\int_{0}^{z}J(z)dz
\ee
which is multi-valued if $J(z)$ has simple poles
(to fix a single-valued branch, we make cuts from
$\infty$ to all simple poles of $J(z)$).
In fact we need only the real part of
$V(z)$.
In neighborhoods of the points $z_{\alpha}$ one has
\be
\label{JV}
J(z)dz=\sum_{k\geq 1}\tau_k dB_k(z),
\;\;\;\;\;
V(z)=\sum_{k\geq 1}\tau_k B_k(z)
\ee
The formula (\ref{n34})
\be\label{defpi}
\Pi_{\alpha} = \phi_\alpha + 2\, {\rm Re}\, V(z_\alpha),
\ \ \ \ \ \ z_\alpha\in{\sf D}_\alpha
\ee
shows that for the algebraic domains the variables
$\Pi_{\alpha}$, introduced in the general case as formal
quantities, are well-defined.
It is easy to show that they are equal to the ${\sf a}$-periods
of the differential $S(z)dz$ on the Schottky double $\Sigma$.
Indeed, using the fact that $C^0(z)$ and $C^{\alpha}(z)$
represent restrictions of the {\em same} function $J(z)$,
one can rewrite (\ref{n201}) in the form
$$
\phi_{\alpha}=-2\, {\rm Re}\left(
\int_{0}^{z_{\alpha}}J(z) dz
+\int_{\xi_0}^{\xi_{\alpha}}(\bar z +C^{-} (z)-J(z))dz \right)
$$
Under the second integral we recognize the Schwarz function.
Combining this equality with the definition of $\Pi_{\alpha}$
(\ref{n34}), we obtain:
\be\label{period33}
\Pi_{\alpha}=2\, {\rm Re}
\int_{\xi_0}^{\xi_{\alpha}} (S(z)-\bar z)dz =
\int_{\xi_0}^{\xi_{\alpha}}
\left (S(z)dz -\bar z d \overline{S(z)} \right )=
\oint_{{\sf a}_{\alpha}} S(z)dz
\ee

As an example of algebraic domains,
it is instructive to consider
the case when only a finite-number of the moments $\tau_k$
are non-zero.
Let ${\cal A}_N$ be the
space of multiply-connected domains such that
\be
\tau_k=0,\ \ \ k>N
\ee
Then the arguments similar to the ones used above show that
\begin{itemize}
\item
{\it S(z) extends to a meromorphic function on $\Sigma$
with a pole
of order $N-1$ at $\infty$ and a simple pole at $\bar{\infty}$}.
\end{itemize}
The function $z$ extended to the lower sheet of the Schottky double
as $\overline{S(z)}$ has a simple pole at $\infty$ and a pole
of order $N-1$ at $\bar{\infty}$.
For a domain ${\sf D^c} \in {\cal A}_N$
the moments with respect to the Laurent
basis (cf. (\ref{momt}))
\be\label{moments}
t_k =-\frac{1}{\pi k}\int_{{\sf D^c}}z^{-k} d^2 z
\ee
coincide with the coefficients of the expansion of the Schwarz
function near $\infty$:
\be\label{n70}
S(z)=\sum_{k=1}^{N}kt_kz^{k-1}+O(z^{-1}),\ \ \ z\to\infty
\ee

The normal displacement of
the bondary of an algebraic domain, which changes the variable $t_k$
keeping all the other moments (and $\Pi_{\alpha}$) fixed
is defined by normal derivative of the function
$2\, {\rm Re}\, \int^z d \tilde \Omega_k$.
Here $d\tilde \Omega_k$ is a unique normalized
meromorphic differential on $\Sigma$  with the only
pole at $\infty$ of the form
\be\label{n60}
d\widetilde \Omega_k=d(z^k+O(z^{-1})),\ \ \ \ \
\oint_{{\sf a}_{\alpha}}d\widetilde \Omega_k=0
\ee
Note that the differential $d\widetilde \Omega_k$ is well-defined for
a generic, not necessarily algebraic domain.
Therefore, the normal derivative
of the function
$2\, {\rm Re}\, \int^z d \tilde \Omega_k$
defines a tangent vector field $\p_{t_k}^{\Pi}$ on the whole space of
multiply-connected domains.

The space
${\cal A}_{N}$ is a particular case of algebraic orbits
of the universal Whitham hierarchy.
In this case the general formula (7.42) from \cite{KriW}
for the $\tau$-function of the Whitham hierarchy,
after proper change of the notation, acquires the form
\be\label{quasihom}
2F=-\frac{1}{2}\tau_{0}^{2} +\tau_0 v_0
+\frac{1}{2} \sum_{k \geq 1}(2-k)
(\tau_k \nu_k +\bar \tau_k \bar \nu_k )-
\sum_{\alpha =1}^{g}\Pi_{\alpha} s_{\alpha}
\ee
which is a quasi-homogeneity condition obeyed by $F$
(compare with formula (5.11) from \cite{MWZ}).

Let $d\Omega_0$ be a unique normalized meromorphic differential on $\Sigma$
with simple poles at the infinities $\infty$ and $\bar\infty$.
Its Abelian integral
\be\label{n71}
\log w(z)=\int^z_{\xi_0} d\Omega_0
\ee
defines in the neighborhood of $\infty$ a function $w(z)$ which
has a simple pole at infinity. The dependence of the inverse function
$z(w)$ on the variables $t_k$ is described by the Whitham
equations for the
two-dimensional Toda lattice hierarchy.
These equations have the form
\be\label{n80}
\p_{t_k}^{\Pi} z(w)=\{\widetilde \Omega_k(w),z(w)\}:=
{d\widetilde\Omega_k(w)\over d\log w}\p_{t_0}z(w)-
\p_{t_0}\widetilde\Omega_k(w){dz\over d\log w}
\ee
Algebraic domains of a more general form
correspond to the universal Whitham hierarchy.
Let ${\cal A}_{N_1,\ldots,N_l}$
be the space of domains such that the extension
of the Schwarz function $S(z)$ to ${\sf D^c}$ has poles of orders
$N_j-1$ at some points $z_j$ (which possibly include $\infty$ and $\bar\infty$).
Then, according to \cite{KriW},
the variables
\be
t_{0,j}=\int^{z_j}_{\xi_0}S(z)dz, \ \
t_{k,j}={1\over k} {\rm res}_{z_j} (z-z_j)^{k-1}S(z)dz,\ \ k=1,\ldots,N_j-1
\ee
together with
the variables $s_{\alpha}$ (or $\Pi_{\alpha}$) provide a set of local
Whitham coordinates on the space ${\cal A}_{N_1,\ldots,N_l}$.
Note that the definition of the algebraic orbits of the universal
Whitham hierarchy is a bit more general than
the definition of algebraic
domains given above.
It corresponds to the case when the differential $dS$ of the Schwarz
function is extendable to ${\sf D^c}$ as a
meromorphic differential (in
\cite{EV} such domains are called Abelian domains).
For example,
let ${\cal A}_N^{(1)}$ be the space of multiply-connected
domains such that
\be
T^{(1)}_k=\oint_{\p{\sf D}}A_{k}dS=\oint_{\p{\sf D}}S'(z)A_{k}dz=
-\oint_{\p {\sf D}}A'_{k}S(z)dz=0, \ \ \ k>N
\ee
This spaces is characterized by the following property:
there are constants $K_{\alpha}$ such that
$S(z)+K_{\alpha}$ extends to a meromorphic function
on ${\sf D^c}$ with a pole
of order $N$ at $\infty$. The variables
\beq
K_{\alpha},\ \ s_{\alpha},\
\ t_k={1\over 2\pi i k}\oint_{\p {\sf D}} z^{-k}\bar zdz ,
\ \ \ k\geq 1,
\eeq
are local coordinates on ${\cal A}_N^{(1)}$.

The two cases when $S(z)$
or its derivative $S'(z)$ have a meromorphic extension
to ${\sf D^c}$ are particular examples of the whole hierarchy of
{\it integrable} domains, which can be defined
in a similar way by the condition that
the $m$-th order derivative of the Schwarz function
$S(z)$ admits a meromorphic extension to the
Schottky double $\Sigma$.
For example, let ${\cal A}_N^{(2)}$
be the space of multiply-connected
domains such that
\be
T^{(2)}_k=\oint_{\p {\sf D}}S''(z)A_{k}dz=
-\oint_{\p {\sf D}}A''_{k}S(z)dz=0,\ \ \ k>N
\ee
This space is characterized by the following property:
there are linear functions
$k_{\alpha}(z)=K^0_{\alpha}+K_{\alpha}^1z$ such that
$S(z)+k_{\alpha}(z)$ extends to a meromorphic function
on ${\sf D^c}$ with a pole
of order $N+1$ at $\infty$. The variables
\be
K_{\alpha}(z),\ \ s_{\alpha},\
\ t_k={1\over 2\pi i k}\oint_{\p {\sf D}} z^{-k}\bar zdz ,
\ \ \ k\geq 1,
\ee
are local coordinates on ${\cal A}_N^{(2)}$. The other spaces
${\cal A}_N^{(m)}$ with $m>2$ can be defined in a similar way.

\section{The duality transformation
\label{ss:dual}}

The independent variables
$\Pi_{\alpha}$ (\ref{period33}) or $\phi_{\alpha}$
used in the previous section
are not as transparent as the
dual variables $s_{\alpha}$ (\ref{dual4}), which
are simply areas of the holes ${\sf D}_{\alpha}$.
In this section we show how to pass
to the set of
independent variables $s_1,\ldots,s_g$ (\ref{dual4})
(together with the infinite set of $\tau_{k}$'s).
This transformation is similar to the passing
from ``external'' to ``internal'' moments
in the simply-connected case (see sect.~5 of \cite{MWZ}).
The difference is that only a finite number of times
are subject to the transformation
while the infinite set of $\tau_k$'s remains the same.

The change to the variables $s_{\alpha}$ can be done in
general case of domains with smooth boundaries. However,
it is the change $\Pi_{\alpha} \to s_{\alpha}$
rather than $\phi_{\alpha} \to s_{\alpha}$ that
leads to a transparent duality. Since $\Pi_{\alpha}$'s
are only defined as formal (``virtual'') variables
for domains with smooth boundaries, we shall restrict
our consideration to the class of algebraic domains
discussed at the end of the previous section.
In this case the variables $\Pi_{\alpha}$ are well defined.

\paragraph{The Legendre transform.}

Passing from $\Pi_{\alpha}$ to $s_{\alpha}$ is a particular
duality transformation which is equivalent to
the interchanging of the ${\sf a}$ and ${\sf b}$ cycles
on the Schottky double $\Sigma$.
This is achieved by the
(partial) Legendre transform
$F(\Pi_{\alpha}, \btau )
\longrightarrow
\tilde F(s_{\alpha}, \btau )$,
where
\be\label{p1}
\tilde F =F +\sum_{\alpha =1}^{g} \Pi_{\alpha}s_{\alpha}
\ee
The function $\tilde F$ is the ``dual'' tau-function.
Below in this section, it is shown that $\tilde F$ solves
the modified Dirichlet problem and can be identified
with the free energy of a
matrix model in the planar
large $N$ limit in the case when the support of eigenvalues
consists of a few disconnected domains (a so called multi-support
solution, see \cite{KM} and references therein).

The main properties of $\tilde F$ follow from those of
$F$. According to (\ref{vk1}), (\ref{dual5}) we have
$dF = -\sum_{\alpha} s_{\alpha}d\Pi_{\alpha}
+\sum_k \nu_k d\tau_k$
(for brevity, $k$ is assumed to run over all integer values,
$\tau_{-k}\equiv \bar \tau_k$, etc.), so
$d\tilde F = \sum_{\alpha} \Pi_{\alpha}ds_{\alpha}
+\sum_k \nu_k d\tau_k$.
This gives the first order derivatives:
\beq\label{f4a}
\Pi_{\alpha}=  \frac{\d \tilde F}{\d s_{\alpha}}\,,
\;\;\;\;
\nu_{k}= \frac{\d \tilde F}{\d \tau_k}
\eeq
The second order derivatives are transformed as follows
(see e.g.\,\cite{dWM}).
Set
$$
F_{\alpha \beta}=\frac{\d^2 F}{\d \Pi_{\alpha} \d \Pi_{\beta}}\,,
\;\;\;
F_{\alpha k}=\frac{\d^2 F}{\d \Pi_{\alpha} \d \tau_k}\,,
\;\;\;
F_{i k}=\frac{\d^2 F}{\d \tau_i \d \tau_k}
$$
and similarly for $\tilde F$.
Then
\beq\label{p2}
\begin{array}{l}
F_{\alpha \beta}=-(\tilde F^{-1})_{\alpha \beta}
\\ \\
\displaystyle{
F_{\alpha k}=\sum_{\gamma =1}^{g}(\tilde F^{-1})_{\alpha \gamma}
\tilde F_{\gamma k} }
\\ \\
\displaystyle{
F_{i k}=\tilde F_{i k} -
\sum_{\gamma , \gamma ' =1}^{g}
\tilde F_{i \gamma }
(\tilde F^{-1})_{\gamma \gamma '}
\tilde F_{\gamma ' k} }
\end{array}
\eeq
Here $(\tilde F^{-1})_{\alpha \beta}$ means the matrix element
of the matrix inverse to the $g\times g$ matrix
$\tilde F_{\alpha \beta}$.

Using these formulas,
it is easy to see that the main properties
(\ref{elem5}), (\ref{elem51}) and (\ref{elem52})
of the tau-function
are translated to the dual tau-function as follows:
\be\label{p3}
\tilde G(a, b )
=\log |a^{-1}-b^{-1}|
+\frac{1}{2} \nabla  (a) \nabla  (b)\tilde F
\ee
\beq\label{p4}
2\pi i \, \tilde \omega_{\alpha}(z)=
-\, \p_{s_{\alpha}}\! \nabla (z) \tilde F
\eeq
\beq\label{p5}
2\pi i \, \tilde T_{\alpha \beta}=
\frac{\p^2 \tilde F}{\p s_{\alpha} \p s_{\beta} }
\eeq
where
$\tau_k$-derivatives in $\nabla (z)$ are taken at
fixed $s_{\alpha}$. The objects in the left hand sides
of these relations are:
\be\label{modified}
\tilde G(a,b)=G(a,b)+i\pi
\sum_{\alpha, \beta =1}^{g}\omega_{\alpha}(a)
\tilde T_{\alpha \beta} \, \omega_{\beta}(b)
\ee
\be\label{modified1}
\tilde \omega_{\alpha}(z)=
\sum_{\beta =1}^{g}\tilde T_{\alpha \beta}\,
\omega_{\beta}(z)
\ee
\be\label{modified2}
\tilde T=-T^{-1} = {i\over\pi}\Omega^{-1}
\ee
The function $\tilde G$ is the Green function of the modified
Dirichlet problem to be discussed below.
The matrix $\tilde T$ is the matrix of ${\sf a}$-periods of the
holomorphic differentials $d\tilde W_\alpha$
on the double $\Sigma$ (so that $\tilde\omega_\alpha(z) = \tilde W_\alpha(z)
+ \overline{\tilde W_\alpha(z)}$), normalized with respect to
the ${\sf b}$-cycles
$$
-\oint_{{\sf b}_\alpha}d\tilde W_\beta = \delta_{\alpha\beta},\ \ \ \ \ \
\oint_{{\sf a}_\alpha}d\tilde W_\beta = \tilde T_{\alpha\beta}
$$
i.e. more precisely, the change of cycles
is ${\sf a}_{\alpha} \to {\sf b}_{\alpha}$,
${\sf b}_{\alpha} \to -{\sf a}_{\alpha}$.

An important remark is in order.
By a simple rescaling of the independent variables
one is able to write
the group of relations (\ref{p3})--(\ref{p5})
for the function $\tilde F$ in exactly the same form as
the ones for the function $F$ (\ref{elem5})--(\ref{elem52}),
so that they differ merely by the notation.
We use this fact in sect.~\ref{ss:eqs}.

\paragraph{The modified Dirichlet problem.}

The modified Green function (\ref{modified})
solves the modified Dirichlet problem which can be formulated in the
following way.

One may eliminate all except for one of the
periods of the Green function $G$, thus making it similar, in this respect, to
the Green function of a simply-connected domain
(recall that the latter has the non-zero period $2\pi$ over the
only boundary curve ${\sf b}_0$).
This leads to the following
{\it modified Dirichlet problem} (see e.g.\, \cite{Gakhov}):
given a function $u_0(z)$ on the boundary,
to find a harmonic function $u(z)$ in
${\sf {\sf D^c}}$ such that it is continuous up
to the boundary and equals
$u_0 (z)+C_{\alpha}$ on the $\alpha$'s
boundary component. Here, $C_{\alpha}$'s are some constants.
It is important to stress that they are not given {\it a priori}
but have to be
determined from the condition that the solution $u(z)$
has vanishing periods over the boundaries
${\sf b}_1,\ldots,{\sf b}_g$.
One of these constants can be put equal to zero.
We set $C_0 =0$.
This problem also has a
unique solution.
It is given by the same formula
(\ref{Dirih}) in terms of the
{\it modified Green function} $\tilde G(z, \zeta )$.
The definition of the latter is similar to that of the
$G(z, \zeta )$ but differs in two respects:
\begin{itemize}
\item[(${\tilde G}1$)]  $\tilde G(z, \zeta )$
is required to have zero periods
over the boundaries ${\sf b}_1,\ldots,{\sf b}_g$;
\item[(${\tilde G}2$)] The derivative of $\tilde G(z, \zeta )$
along the boundary
(not $\tilde G(z, \zeta )$ itself!) vanishes on the boundary.
\end{itemize}
Under the condition that
$\tilde G(z, \zeta )=0$ on ${\sf b}_0$ such a function is unique.
The function given by (\ref{modified}) just meets
these requirements.
We conclude that the modified
Green function is expressed through the dual tau-function $\tilde F$.

Note that variations of the modified Green function
under small deformations of the domain are described by
the same Hadamard formula (\ref{Hadam}), where each Green
function is replaced by $\tilde G$.
This follows, after some algebra, from the
formula for $\tilde G$ in terms of $G$,
$\omega_{\alpha}$ and $\Omega_{\alpha \beta}$.
Therefore, all the
arguments of sect.~\ref{ss:taumu} could be
repeated in a completely parallel way starting from the modified
Dirichlet problem.
One may also say that the functions $G$ and $\tilde G$
differ merely by a prefered basis of cycles on the double:
the differential
$\d_z G dz$ has vanishing periods over the ${\sf a}$-cycles
while $\d_z \tilde G dz$ has vanishing periods over
the ${\sf b}$-cycles.

\begin{figure}[tb]
%\hspace*{2cm}
\epsfysize=5cm
\centerline{\epsfbox{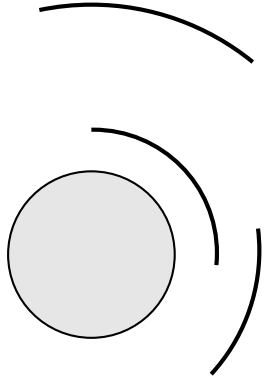}}
%\centerline{\epsfbox{extcut.eps}}
\caption{\sl
The image
of a triply-connected domain
under the conformal map $w_a(z)$
such that $\tilde G(z,a )= -\log |w_a (z)|$.
The function $w_a$ maps the domain onto
the exterior of the unit (dashed) circle
with $g=3$ concentric circular cuts.
Positions and lengths of the arcs
depend on the shape of
${\sf {\sf D^c}}$ in fig.~\ref{fi:multid} and
depend also on the normalization point
$z=a$ which is mapped to $\infty$.}
\label{fi:extcut}
\end{figure}

The function $w_a (z)$ such that
$\tilde G(z,a )= -\log |w_a (z)|$
maps ${\sf D^c}$ onto the exterior of the unit
circle which is slit along $g$ concentric circular arcs
(see fig.~\ref{fi:extcut}). Since the periods
of the function $\tilde G$ vanish, the function $w_a$ is
single-valued.
Positions of the arcs depend on the shape of ${\sf D^c}$ as well as
on the point $a$ which is mapped to $\infty$.
The radii of the arcs, $R_{\alpha}$, are expressed through
the dual tau-function as
\beq\label{radii}
\log R_{\alpha}^{2}=\p_{s_{\alpha}}\! \nabla (a)\tilde F
\eeq
In particular, for $a=\infty$ we have
$\log R_{\alpha}^{2}=\p_{s_{\alpha}}\! \p_{\tau_0}\tilde F$
(cf. eq.\,(\ref{sec7}) for the conformal radius).

\paragraph{Relation to multi-support solutions
of matrix models.}
The partition function of the model\footnote{One may
have in mind the model of all complex or mutually hermitean
conjugated or normal matrices.},
written as an integral over
eigenvalues, reads:
\be\label{mat1}
Z_N =\frac{1}{N!}\int \exp \left ( \sum_{i<j}^{N}
\log |z_i -z_j |^2 +\frac{1}{\hbar}\sum_{i=1}^{N} U(z_i)\right )
\prod_{j=1}^{N}d^2 z_j
\ee
The matrix model potential $U$ is usually chosen to be of the form
\be\label{mat2}
U(z)=-z\bar z  +V(z)+\overline{V(z)}
\ee
where $V(z)$ is at the moment some polynomial; however, we will see
immediately that the coincidence of the notation with (\ref{JV}) is not accidental.
The parameter $\hbar$, in the
large $N$ limit, tends to zero
simultaneously with $N \to \infty$
in such a way that $t_0 =N\hbar $ is kept finite and fixed.
In the leading order, one can apply the saddle point method.
The saddle point condition is
$$
\hbar \!\sum_{j=1, \neq i}^{N}\frac{1}{z_i -z_j}
+\p_{z_i}U(z_i) =0
$$
The density of eigenvalues is a sum of two-dimensional
delta-functions:
$$
\rho (z)=\pi\hbar \sum_i \delta^{(2)}(z-z_i)
$$
In the large $N$ limit, one treats it
as a continuous function normalized as ${1\over\pi}\int \rho (z)d^2z=t_0$.
In terms of the continuous density the saddle point equation reads
\be\label{semitau1}
{1\over\pi}\int \frac{ \rho (\zeta ) d^2\zeta}{z-\zeta}
+\p_z U(z) =0
\ee
The solution is well known and easy to obtain:
the extremal density $\rho_0 (z)$
is constant in some domain ${\sf D}$
(the support of eigenvalues which
can be a union of disconnected domains ${\sf D}_{\alpha}$) and zero
otherwise. More precisely,
$$
\rho_0 (z)=\left \{
\begin{array}{ll}
1\;\; & z\in {\sf D}
\\
0\;\; & z\in {\bf C}\setminus {\sf D} \equiv {\sf D^c}
\end{array}\right.
$$
Note that the saddle point condition is imposed
only for $z$ inside the support of eigenvalues.
Writing (cf. (\ref{JV}))
$$
V(z)=\sum_{k=1}^{p}\tau_k B_k (z)
$$
multiplying eq.\,(\ref{semitau1}) by $A_k (z)$ and integrating
it over the boundary of the support of eigenvalues, one finds
that the domain ${\sf D}$ is such that
the coefficients $\tau_k$ are moments of its complement
with respect to the basis functions $A_k$ and the higher moments
(with numbers greater than $p$) vanish. As is proven in sect.~\ref{ss:taumu},
these conditions, together with the normalization condition for the
density, locally determine the shape of the support of eigenvalues.
Equivalently, one might parametrize the polynomial as
$V(z)=\sum_{k=1}^{p}t_k z^{k}$, then $t_k$ are moments
of the ${\bf C}\setminus {\sf D}$ with respect to the functions $z^{-k}$.

An interesting problem is to obtain, for a given value of $t_0$,
necessary and
sufficient conditions on the
polynomial $V$ for the support of eigenvalues to be a union
of $g+1$ disconnected domains (''droplets'') with non-zero filling.
One may approach this problem from a ``classical'' limit
of very small (point-like) droplets. Clearly, for our choice
of the signs in (\ref{mat1}) and (\ref{mat2}), the stable point-like
droplets are located at minima of $- U(z)$, or equivalently at maxima
of $U(z)$. As soon as we use the basis which
explicitly depends on the marked points $z_{\alpha}\in {\sf D}_{\alpha}$,
it is
natural to consider the germ configuration with point-like
eigenvalue droplets at the points $z_{\alpha}$. It is easy
to see that the sufficient conditions for the potential (\ref{mat2})
to have maxima at the points $z_{\alpha}$ are
\beq\label{PP}
\bar z_{\alpha} = V'(z_{\alpha}),
\;\;\;\;
|V''(z_{\alpha})|<1
\;\;\;
\mbox{for all $\alpha$}
\eeq
The first one means that there is
an extremum of the potential $U(z)$ at the point $z_{\alpha}$.
The second one ensures that eigenvalues of the matrix of
second derivatives of the potential at the extremum are
both negative, i.e. the extremum is actually a maximum
of $U(z)$. The first condition literally coincides with the one used in \cite{KM}
for the completely degenerate curve (point-like droplets) while the second one requires that
not all extrema are filled, so that the ``smooth" genus $g$ is always
less than the maximal possible genus $(p-1)^2-1$.
Let  now $V_0 (z)$ be a minimal degree polynomial that
obeys these conditions. Then perturbed polynomials of the form
$$
V(z)=V_0(z)+\sum_{k\geq g+2}\tau_k B_k (z)
$$
obey the same conditions for sufficiently small $\tau_k$, and this is
the advantage of using the basis (\ref{a}), (\ref{o}) from the
perspective of matrix models.

The saddle point equation (\ref{semitau1}) just means that the ``effective
potential" $\frac{1}{\pi}\int_{\sf D}
\log |z-\zeta |^2 d^2 \zeta +U(z)$ is constant
in each ${\sf D}_{\alpha}$, i.e. for $z\in {\sf D}_{\alpha}$
it holds:
\be\label{constpi}
\frac{2}{\pi}\int_{\sf D} \log |z-\zeta | d^2 \zeta +U(z)=
v_0 +\Pi_{\alpha}
\ee
where
$v_0 =\frac{2}{\pi}\int_{\sf D} \log |z| d^2 z$,
and so in our normalization $\Pi_0 =0$.

Let $N_{\alpha}$ be the number of eigenvalues in ${\sf D}_{\alpha}$,
then $\lim _{N\to \infty}\hbar N_{\alpha}=s_{\alpha}$.
First we find the free energy with some fixed $s_{\alpha}$:
$$
\left. \lim_{N\to \infty}
\left (\hbar^2 \log Z_N \right )
\right |_{s_{\alpha} \; {\rm fixed}}=
{1\over\pi^2}\int \!\int \rho_0 (z)\log |z-\zeta |\rho_0 (\zeta)
d^2 z d^2 \zeta +{1\over\pi}\int \rho_0 (z)U(z)d^2z
$$
where one should substitute (\ref{constpi}).
The result is
\beq\label{mat5}
\left. \lim_{N\to \infty}
\left (\hbar^2 \log Z_N \right )
\right |_{s_{\alpha} \; {\rm fixed}} \! =
-\,\frac{1}{\pi^2}\int_{{\sf D}} \!\int_{{\sf D}}
\log |z^{-1}-\zeta^{-1}|d^2 z d^2 \zeta +
\sum_{\alpha =1}^{g}\Pi_{\alpha}s_{\alpha}=
\tilde F(s_{\alpha}, \btau )
\eeq
This quantity depends on $s_{\alpha}$'s. It is given by the
value of the integrand in (\ref{mat1}) at the saddle point
with fixed $s_{\alpha}$.

If one wants to take into account the ``tunneling''
of eigenvalues
between different components of the support,
$s_{\alpha}$ are no longer free parameters, and
the free energy $F^{(0)}$ of the ``planar limit" of the matrix model
should be obtained by extremizing $\tilde F$ with respect to
$s_{\alpha}$ (with fixed $t_0$). It follows from the above that
$\p_{s_{\alpha}}\tilde F=\Pi_{\alpha}$
and so the extremum is at $\Pi_{\alpha}=0$.
Therefore,
\be\label{mat6}
\left. \phantom{A/B}
F^{(0)}= F\right |_{\Pi_{\alpha}=0}
\ee
where $F$ is given by (\ref{F}). The results of sect.~\ref{ss:taumu}
imply that the growth of the support of eigenvalues,
when $N \to N+\delta N$ and the tunneling
is taken into account, is given by the normal displacement of the boundary
proportional to the normal derivative of the Green
function with the minus sign,
like in (\ref{small}). Since this quantity is
always non-negative, we conclude that
if a point belongs to the support of eigenvalues,
it does so as $t_0$ increases.
In particular, if one starts with point-like droplets
at the points $z_{\alpha}$, as is discussed above, then these points
always remain inside the droplets.

Let us note that different aspects
of multi-support solutions of
the 2-matrix model and matrix models with complex eigenvalues
were discussed in \cite{KM,Bertola,Eynard}.

For matrix models one usually restricts
oneself to the algebraic case of
finite amount of nonvanishing moments playing the role
of the coefficients of the matrix model potential (\ref{mat2}). In such case the
corresponding complex curve can be described by an algebraic equation
\cite{KM}, which can be thought of
as auxiliary constraint to the second
derivatives of $F$. These auxiliary constraints look similar to the
reduction conditions in the case of Landau-Ginzburg topological theories
(see, for example, discussion of such conditions in the context of
dispersionless Hirota equations and WDVV equations in \cite{BMRWZ}).
Their
meaning is that the derivatives w.r.t. the times $\{ \tau_k \}$ for
$k>g+1$ (where $g$ is genus
of corresponding algebraic complex curve) can be expressed through the
derivatives restricted to $k\leq g+1$.

\section{Green function on the Schottky double and
generalized Hirota equations
\label{ss:eqs}}

Let us now turn to the generalization of the dispersionless
Hirota equations to the multiply-connected case.

For a unified treatment of the two ``dual'' representations
of the Dirichlet problem discussed in the two previous sections,
we make a simple change of variables. Namely, let us
introduce the generic ``period" variables $X_{\alpha}$ which are
identified either with
$\Pi_{\alpha}$ or $2\pi i \, s_{\alpha}$ depending on the
choice of the set of cycles,
and the
function ${\cal F}(X_{\alpha}, \btau )$ equal to
$F(X_{\alpha}, \btau )$ or
$\tilde F(\frac{X_{\alpha}}{2\pi i}, \btau )$ respectively.
Then the main relations (\ref{elem5}) -- (\ref{elem52})
and (\ref{p3}) -- (\ref{p5}) acquire the form
\be\label{u3}
G(a, b )
=\log |a^{-1}-b^{-1}|
+\frac{1}{2} \nabla  (a) \nabla  (b) {\cal F}
\ee
\beq\label{u4}
\omega_{\alpha}(z)=
-\, \p_{\alpha} \nabla (z) {\cal F}
\eeq
\beq\label{u5}
T_{\alpha \beta}=
2\pi i \,
\p_{\alpha} \p_{\beta} {\cal F}
\eeq
where $\p_{\alpha}:=\p_{X_{\alpha}}$
and $G$, $\omega$ and $T$ stand for the
corresponding objects
with or without tilde, depending on the chosen basis
of cycles.

We will see below that,
in analogy to the simply-conected case,
any second order derivative of the function
${\cal F}$ w.r.t. $\tau_k$ (and $\bar \tau_k$), ${\cal F}_{ik}$,
will be expressed through the
derivatives
$\{ {\cal F}_{\alpha\beta}\}$ where
$\alpha,\beta=0,\ldots,g$ together
with $\{ {\cal F}_{\alpha \tau_i}\}$ and their complex conjugated.
To be more
precise, one can consider all second derivatives as functions of
$\{ {\cal F}_{\alpha\beta}, {\cal F}_{\alpha k} \}$ modulo
certain relations on
the latter, like the relation (\ref{corad}) to be
discussed below. Sometimes on this
``small phase space" more extra
constraints arise, which can be written in the form similar to the
Hirota or WDVV equations \cite{MMMBr};
we are not going to discuss this
issue here, restricting ourselves to the generic situation.

\paragraph{The Abel map.}
To derive equations for the function $F={\cal F}$
in sect.~\ref{ss:simply},
we used the representation (\ref{sec4})
for the conformal map $w(z)$ in terms of ${\cal F}$ and
eq.\,(\ref{Gconf}) relating the conformal map to the
Green function which, in its turn, is expressed
through the second order derivatives of ${\cal F}$.
In the multiply-connected case,
our strategy is basically the same, with
the suitable analog of the conformal map
$w(z)$ (or rather of $\log w(z)$)
being the embedding of ${\sf D^c}$ into the
$g$-dimensional complex torus ${\bf Jac}$,
the Jacobi variety of the Schottky double.
This embedding is given, up to an overall
shift in ${\bf Jac}$, by the Abel map
$z \mapsto {\bf W}(z):= (W_1 (z), \ldots , W_g (z) )$
where
\beq\label{E1}
W_{\alpha}(z)=\int_{\xi_0}^{z} dW_{\alpha}
\eeq
is the holomorphic part of the
harmonic measure $\omega_{\alpha}$.
By virtue of (\ref{u4}), the Abel map
is represented through the second order derivatives
of the function ${\cal F}$:
\beq\label{E2}
W_{\alpha}(z)-W_{\alpha}(\infty)=
\int_{\infty}^{z} dW_{\alpha}=
-\p_{\alpha}D(z){\cal F}
\eeq
\beq\label{E3}
2\, {\rm Re}\, W_{\alpha}(\infty)=
\omega_{\alpha}(\infty)=-\p_{\tau_0}\p_{\alpha}{\cal F}
\eeq
The last formula immediately follows from (\ref{u4}).

\paragraph{The Green function and the prime form.}
The Green function of the Dirichlet
boundary problem, appearing
in (\ref{u3}), can be written in terms of the prime form
(\ref{primed})
on the Schottky double (cf. (\ref{Gconf})):
\be
\label{prime}
G(z,\zeta) = \log\left|E(z,\zeta)\over E(z, \bar\zeta)\right|
\ee
Here by $\bar\zeta$ we mean the (holomorphic) coordinate of
the ``mirror" point on the Schottky double, i.e.
the ``mirror" of $\zeta$ under the
antiholomorphic involution. The pairs of such mirror
points satisfy the condition
$W_\alpha(\zeta ) + \overline{W_\alpha( \bar \zeta)} = 0$
in the Jacobian (i.e., the sum should be
zero modulo the lattice of periods).
The prime form\footnote{Given a Riemann surface with local
coordinates $1/z$ and
$1/\bar z$ we trivialize the bundle of $-\ha$-differentials and
``redefine" the prime form $E(z,\zeta)\rightarrow
E(z,\zeta)(dz)^{1/2}(d\zeta)^{1/2}$ so that it becomes a function.
However for different coordinate patches
(the ``upper" and ``lower" sheets
of the Schottky double) one gets
different functions, see,
for example, formulas (\ref{Prime1}) and (\ref{Prime2}) below.}
is written through the Riemann theta functions
and the Abel map as follows:
\beq\label{Prime1}
E(z, \zeta )=\frac{\theta_\ast
({\bf W}(z)-{\bf W}(\zeta))}{h(z)\, h(\zeta )}
\eeq
when the both points are on the upper sheet and
\beq\label{Prime2}
E(z, \bar \zeta )=\frac{\theta_\ast
({\bf W}(z)+\overline{{\bf W}(\zeta)})}{ih(z)\,
\overline{h(\zeta )}}
\eeq
when $z$ is on the upper sheet and
$\bar \zeta$ is on the lower one
(for other cases we define
$E(\bar z, \bar \zeta )=\overline{E(z, \zeta )}$,
$E(\bar z, \zeta )=\overline{E(z, \bar \zeta )}$).
Here $\theta_\ast ({\bf W})
\equiv \theta_{\bdelta^\ast} ({\bf W} | T)$ is
the Riemann theta function (\ref{thetad})
with the period matrix $T_{\alpha \beta}=
2\pi i \, \p_{\alpha} \p_{\beta}{\cal F}$
and any odd characteristics $\bdelta^\ast$, and
\beq\label{Prime3}
h^2 (z)=-z^2 \sum_{\alpha =1}^{g}\theta_{\ast , \alpha}(0)
\p_z W_{\alpha}(z)
=z^2 \sum_{\alpha =1}^{g}\theta_{\ast , \alpha}(0)
\sum_{k\geq 1}A_{k}'(z)\p_{\alpha}\p_{\tau_k}{\cal F}
\eeq
Note that in the l.h.s. of (\ref{Prime2}) the bar
means the reflection in the double while
in the r.h.s. the bar means complex conjugation.
The notation is consistent since the local coordinate
in the lower sheet is just the complex conjugate one.
However, one should remember that $E(z, \bar \zeta)$
is {\it not} obtained from (\ref{Prime1}) by a simple
substitution of the complex conjugated argument.
On different sheets so defined
prime ``form" $E$ is represented by different
functions.
In our normalization (\ref{Prime2})
$iE(z, \bar z)$ is real (see also Appendix B) and
$$
\lim_{\zeta \to z}\frac{E(z,\zeta )}{z^{-1}-\zeta^{-1}} =1
$$
In particular, $\lim_{z\to \infty} zE(z, \infty )=1$.

\paragraph{The prime form and the tau-function.}
In (\ref{prime}), the $h$-functions in the
prime forms cancel, so the analog of
(\ref{Gconf1}) reads
\be
\label{gfm}
\log\left|\theta_\ast ({\bf W}(z)-{\bf W}(\zeta))
\over \theta_\ast ({\bf W}(z)+
\overline{\bf W(\zeta)})\right|^2=
\log\left|{1\over z} - {1\over \zeta }\right|^2
+\nabla(z)\nabla(\zeta ){\cal F}
\ee
This equation already explains the claim made in
the beginning of this section.
Indeed, the r.h.s. is the generating function for the
derivatives ${\cal F}_{ik}$ while the l.h.s. is expressed through
derivatives of the form ${\cal F}_{\alpha k}$
and ${\cal F}_{\alpha \beta}$
only. The expansion in powers of $z, \zeta$ allows one
to express the former through the latter.

The analogs of eqs.\,(\ref{sec3a}), (\ref{sec7})
are, respectively:
\beq\label{gfm1}
\log\left|\theta_\ast ({\bf W}(z)-{\bf W}(\infty))
\over \theta_\ast ({\bf W}(z)+\overline{{\bf W}(\infty )})\right|^2 =
-\log|z|^2 +\p_{\tau_0} \nabla (z){\cal F}
\eeq
\beq\label{gfm2}
\log \left |\frac{h^{2}(\infty)}{\theta_{\ast}
(\bomega (\infty ))}\right |^2 =\p_{\tau_0}^{2}{\cal F}
\eeq
Here $\bomega (z)\equiv 2\, {\rm Re}\,\bW (z)=
(\omega_1 (z), \ldots , \omega_g (z))$ and
$$
h^2 (\infty)=\lim_{z\to \infty}z\, \theta_{\ast}
\left ( \int_{\infty}^{z}d\bW \right )=
-\sum_{\alpha =1}^{g}\theta_{\ast , \alpha}(0)
\p_{\alpha}\p_{\tau_1}{\cal F}
$$
A simple check shows that the
l.h.s. of (\ref{gfm2}) can be written as
$-2\log (iE (\infty , \bar \infty ))$.
As is seen from the expansion
$G(z, \infty )=-\log |z| -
\log (i E(\infty , \bar \infty )) + O(z^{-1}) $ as
$z\to \infty$, $(iE (\infty , \bar \infty ))^{-1}$ is
a natural analog of the conformal radius, and
(\ref{gfm2}) indeed turns to (\ref{sec7}) in the
simply-connected case (see Appendix B for an explicit
illustrative example).
However, now it provides
a nontrivial relation on ${\cal F}_{\alpha\beta}$'s and
$ {\cal F}_{\alpha i}$'s:
\be
\label{corad}
\left (\sum_{\alpha}\theta_{\ast , \alpha}
\p_{\alpha}\p_{\tau_1}{\cal F}\right )
\left (\sum_{\beta}\theta_{\ast , \beta}
\p_{\beta}\p_{\bar \tau_1}{\cal F}\right )
=\theta^{2}_{\ast}
(\bomega (\infty )) e^{\p_{\tau_0}^{2}{\cal F}}
\ee
so that the ``small phase space" contains the
derivatives modulo this relation.

The next
steps are exactly the same as in sect.~\ref{ss:simply}:
we are going to decompose these equalities into
holomorphic and antiholomorphic parts.
The results are conveniently written in terms of
the prime form. The counterpart of (\ref{sec6}) is
\beq\label{analsec6}
\log \frac{E(\zeta , z)E(\infty ,
\bar \zeta )}{E(\zeta , \infty )E(z, \bar \zeta )}=
\log \left ( 1-\frac{\zeta}{z}\right )
+D(z) \nabla (\zeta ){\cal F}
\eeq
Tending $\zeta \to \infty$, we get:
\beq\label{analsec4}
\log \frac{E(z, \bar \infty )}{E(z, \infty )}=
\log z +\log E(\infty , \bar \infty )-
\p_{\tau_0}D(z){\cal F}
\eeq
Separating holomorphic and antiholomorphic
parts of (\ref{analsec6}) in $\zeta$, we get
analogs of (\ref{sec5}) and (\ref{511}):
\beq\label{analsec5}
\log \frac{E(z, \zeta )}{E(z, \infty )E(\infty , \zeta)}
=\log (z-\zeta) +D(z)D(\zeta){\cal F}
\eeq
\beq\label{anal511}
-\log \frac{E(z, \bar \zeta )
E(\infty , \bar \infty )}{E(z, \bar \infty )
E(\infty , \bar \zeta)} =D(z)\bar D(\bar \zeta){\cal F}
\eeq
Combining these equalities (with merging points
$z\to \zeta$ in particular), one is able to
obtain the following
representations of the prime form itself:
\beq\label{E10}
E(z, \zeta )=(z^{-1}-\zeta^{-1})
e^{-\frac{1}{2}(D(z)-D(\zeta ))^2 {\cal F}}
\eeq
\beq\label{E11}
iE(z, \bar \zeta )=
e^{-\frac{1}{2}(\p_{\tau_0} +D(z)+ \bar D(\bar \zeta ))^2 {\cal F}}
\eeq
Note also the nice formula
\beq\label{E12}
iE (z, \bar z )=
e^{-\frac{1}{2}\nabla^2 (z) {\cal F}}
\eeq

\paragraph{Generalized Hirota relations.}
For higher genus Riemann surfaces there are no
simple universal relations connecting values of
prime forms at different points, which,
via (\ref{E10}), (\ref{E11}), could be used
to generate equations on ${\cal F}$.
The best available relation \cite{Fay} is the
celebrated Fay identity (\ref{fay1}). Although
it contains not only prime forms but Riemann theta
functions themselves, it is really a source
of closed equations on ${\cal F}$, since all the ingredients
are in fact representable in terms of
second order derivatives of ${\cal F}$ in different variables.

An analog of the KP version of the Hirota equation (\ref{Hir1})
for the function ${\cal F}$
can be obtained
by plugging eqs.\,(\ref{E2}) and
(\ref{E10}) into
the Fay identity (\ref{fay1}). As a result, one obtains
a closed equation which contains second order derivatives
of the ${\cal F}$ only (recall that the period matrix
in the theta-functions is essentially the matrix
of the derivatives ${\cal F}_{\alpha \beta}$).
A few equivalent forms of this equation are available.
First, shifting
$\bZ\rightarrow\bZ-\bW_3+\bW_4$ in (\ref{fay1}) and
putting $z_4 =\infty$, one gets the relation
\be
\label{HirKPsym}
(a-b)e^{D(a)D(b){\cal F}}\
\theta \left (\int_{\infty}^{a}d\bW + \int_{\infty}^{b}d\bW -\bZ \right )
\ \theta \left (\int_{\infty}^{c}d\bW -\bZ \right )\, +
\\ +  \,
(b-c)e^{D(b)D(c){\cal F}}\
\theta \left (\int_{\infty}^{b}d\bW + \int_{\infty}^{c}d\bW -\bZ \right )
\ \theta \left (\int_{\infty}^{a}d\bW -\bZ \right )\, +
\\ +\,
(c-a)e^{D(c)D(a){\cal F}}\
\theta \left (\int_{\infty}^{c}d\bW + \int_{\infty}^{a}d\bW -\bZ \right )
\ \theta \left (\int_{\infty}^{b}d\bW -\bZ \right )\, =0
\ee
The vector $\bZ$ is arbitrary (in particular, zero).
We see that (\ref{Hir1}) gets ``dressed'' by the theta-factors.
Each theta-factor is expressed through ${\cal F}$ only.
For example,
$$
\theta \left (\int_{\infty}^{z}d\bW  \right )=
\sum_{n_{\alpha}\in \bZ}\exp \left (
 -2\pi^2 \sum_{\alpha \beta}n_{\alpha}n_{\beta}
\p_{\alpha \beta}^{2}{\cal F} -
2\pi i \sum_{\alpha}n_{\alpha}
\p_{\alpha}D(z){\cal F}\right )
$$
Another form of this equation, obtained from
(\ref{HirKPsym}) for a particular choice of $\bZ$, reads
\beq\label{HirKPsym1}
(a\!-\! b)c^{-1}e^{\frac{1}{2}(D(a)+D(b))^2 {\cal F}}
h(c)
\theta_{\ast}
\left (\int_{\infty}^{a}d\bW \!+\! \int_{\infty}^{b}d\bW \right ) +
[\mbox{cyclic per-s of $a,b,c$}] =0
\eeq
Taking the limit
$c\to \infty$ in (\ref{HirKPsym}), one gets
an analog of (\ref{Hir2}):
\be
\label{HirKP2}
1 \, - \,
{\theta (\int_{\infty}^{a}d\bW +\int_{\infty}^{b}
d\bW-\bZ)\ \ \theta (\bZ)\over
\theta (\int_{\infty}^{a}d\bW -\bZ)\
\theta ( \int_{\infty}^{b}d\bW-\bZ)}
\, e^{D(a)D(b){\cal F}} \, =
\\
= \, \frac{D(a)-D(b)}{a-b}\d_{\tau_1}{\cal F}
+\frac{1}{a-b}\, \sum_{\alpha=1}^g
\frac{\p}{\p Z_{\alpha}}
\log {\theta (\int_{\infty}^{a} d\bW
-\bZ)\over \theta (\int_{\infty}^{b}\bW -\bZ)}
\, \p_{\alpha}  \p_{\tau_1}{\cal F}
\ee
which also follows from another Fay identity (\ref{fay2}).

Equations on ${\cal F}$ with $\bar \tau_k$-derivatives follow
from the general Fay identity (\ref{fay1}) with
some points on the lower sheet.
Besides, many other equations can be derived as various
combinations and specializations of the ones mentioned
above. Altogether, they form an infinite hierarchy of
consistent differential equations of a very complicated
structure which deserves further investigation.
The functions ${\cal F}$ corresponding to different choices
of independent variables (i.e., to different bases in
homology cycles on the Schottky double) provide different
solutions to this hierarchy.

\paragraph{Higher genus analogs of the dispersionless Toda
equation.}
Let us show how the simplest equation of the hierarchy,
the dispersionless Toda equation (\ref{Toda}),
is modified in the multiply-connected case.
Applying $\p_z \p_{\bar \zeta}$
to both sides of (\ref{anal511}) and setting
$\zeta =z$, we get:
$$
(\p D(z))(\bar \p \bar D(\bar z)){\cal F} =
-\p_z \p_{\bar z}\log E(z, \bar z)
$$
Here $\p D(z)$ is the $z$-derivative of the operator
$D(z)$: $\p D(z)=\sum_k A'_{k}(z)\p_{\tau_k}$.
To transform the r.h.s.,
we use the identity (\ref{fay3})
(Appendix A) and
specialize it to the particular local parameters
on the two sheets:
$$
|z|^4 \p_z \p_{\bar z}\log E(z, \bar z)=
\frac{\theta (\bomega (z) +\bZ )
\theta (\bomega (z)-\bZ )}{\theta^2 (\bZ )
E^2 (z,\bar z)} +
|z|^4 \sum_{\alpha, \beta}(\log
\theta ({\bf Z}))_{, \, \alpha \beta}
\, \p_z W_{\alpha}(z)\p_{\bar z}
\overline{W_{\beta}(z)}
$$
Tending $z$ to $\infty$,
we obtain a family of equations
(parametrized by an arbitrary vector $\bZ$) which
generalize the dispesrionless Toda equation for the
tau-function:
\beq\label{Todagen}
\p_{\tau_1} \p_{\bar \tau_1} {\cal F}
=\frac{\theta (\bomega (\infty ) \!+\!\bZ )
\theta (\bomega (\infty )\!-\! \bZ )}{\theta^2 (\bZ )}
\, e^{\p_{\tau_0}^{2}{\cal F}}\,
-\!\! \sum_{\alpha , \beta =1}^{g}
\! (\log \theta (\bZ ))_{, \, \alpha \beta}
\, (\p_{\alpha} \p_{\tau_1}{\cal F})
(\p_{\beta} \p_{\bar \tau_1}{\cal F})
\eeq
(Here we used the $z\to \infty$ limits
of (\ref{E2}) and (\ref{E12}).)
The following two equations correspond to
special choices of the vector $\bZ$:
\be\label{Toda2}
\p_{\tau_1} \p_{\bar \tau_1} {\cal F}
+ \sum_{\alpha , \beta =1}^{g}
(\log \theta (0))_{, \, \alpha \beta}
\, (\p_{\alpha} \p_{\tau_1}{\cal F})
(\p_{\beta} \p_{\bar \tau_1}{\cal F})
=\frac{\theta^2 (\bomega (\infty ))}{\theta^2 (0)}
\ e^{\p_{\tau_0}^{2}{\cal F}}\,
\ee
\be
\label{Toda1}
\p_{\tau_1} \p_{\bar \tau_1} {\cal F}
=- \sum_{\alpha,\beta =1}^{g}
\left[\log\theta_{\ast}(\bomega (\infty))\right]_{,\alpha\beta}
(\p_{\alpha} \p_{\tau_1}{\cal F})
(\p_{\beta} \p_{\bar \tau_1}{\cal F})
\ee

Finally, let us specify the equation (\ref{Todagen})
for the genus $g=1$ case.
In this case there is only one extra
variable
$X :=X_1$
($\Pi_1$ or $2\pi is_1$).
The Riemann theta-function
$\theta (\bomega (\infty ) +\bZ )$
is then replaced by the Jacobi theta-function
$\vartheta \left (\left. \p_{X}\p_{\tau_0}{\cal F}
 -Z \right |T \right )
\equiv \vartheta_3 \left (\left. \p_{X}\p_{\tau_0}{\cal F}
 -Z \right |T \right )$,
where the elliptic modular parameter is
$T =2\pi i \, \p_X^2 {\cal F}$,
and the vector $Z\equiv {\bf Z}$ has only one component.
The equation has the form:
\beq\label{Todagen1}
\begin{array}{c}
\displaystyle{
\p_{\tau_1} \p_{\bar \tau_1} {\cal F}
=\frac{\vartheta_3 \left (\left. \p_{X}\p_{\tau_0}{\cal F}
\! +\! Z \right |  2\pi i \, \p_X^2 {\cal F} \right )
\vartheta_3 \left (\left. \p_{X}\p_{\tau_0}{\cal F}
 \!-\! Z \right |2\pi i \, \p_X^2 {\cal F}\right )}{\vartheta_3^2
\left (\left. Z \right |2\pi i \, \p_X^2 {\cal F} \right )}
\, e^{\p_{\tau_0}^{2}{\cal F}} }
\\ \\
\displaystyle{-\,\,  \p_{Z}^{2}\log \vartheta_3
\left (\left. Z \right |2\pi i \, \p_X^2 {\cal F} \right )
\, (\p_X \p_{\tau_1}{\cal F})
(\p_X \p_{\bar \tau_1}{\cal F})}
\end{array}
\eeq
Note also that equation (\ref{corad}) acquires the form
\beq\label{corad1}
(\p_X \p_{\tau_1}{\cal F})
(\p_X \p_{\bar \tau_1}{\cal F})=
\left (
\frac{\vartheta_1 \left (\left. \p_{X}\p_{\tau_0}{\cal F}
\right |  2\pi i \,
\p_X^2 {\cal F} \right )}{\vartheta_{1}'
\left (\left. 0 \right |  2\pi i \, \p_X^2 {\cal F} \right )}
\right )^2
e^{\p_{\tau_0}^{2}{\cal F}}
\eeq
where $\vartheta_\ast\equiv\vartheta_1$ is the only
odd Jacobi theta-function. Combining (\ref{Todagen1}) and (\ref{corad1})
one may also write the equation
\be\label{Todagell}
\p_{\tau_1} \p_{\bar \tau_1} {\cal F}
= \left[
\frac{\vartheta_3 \left (\left. \p_{X}\p_{\tau_0}{\cal F}
\! +\! Z \right |  2\pi i \, \p_X^2 {\cal F} \right )
\vartheta_3 \left (\left. \p_{X}\p_{\tau_0}{\cal F}
 \!-\! Z \right |2\pi i \, \p_X^2 {\cal F}\right )}{\vartheta_3^2
\left (\left. Z \right |2\pi i \, \p_X^2 {\cal F} \right )}
\right.
\\
\left.-\,\,
\left (
\frac{\vartheta_1 \left (\left. \p_{X}\p_{\tau_0}{\cal F}
\right |  2\pi i \,
\p_X^2 {\cal F} \right )}{\vartheta_{1}'
\left (\left. 0 \right |  2\pi i \, \p_X^2 {\cal F} \right )}
\right )^2
\p_{Z}^{2}\log \vartheta_3
\left (\left. Z \right |2\pi i \, \p_X^2 {\cal F} \right )
\right] \, e^{\p_{\tau_0}^{2}{\cal F}}\,
\ee
whose form is close to (\ref{Toda}) but
differs by the nontrivial
``coefficient" in the square brackets. In the
limit $T \to i\infty$ the theta-function $\vartheta_3$
tends to unity, and we obtain the dispersionless
Toda equation (\ref{Toda}).

\section{Conclusions}

In this paper we have considered
the Dirichlet boundary problem in
planar multiply-connected domains.
A planar multiply-connected domain ${\sf D^c}$ is the complex
plane with several holes.
We study how the solution of the Dirichlet problem
depends on small deformations of boundaries of the holes.

General properties of such deformations
allow us to introduce the quasiclassical tau-function
associated to the variety of planar multiply-connected domains.
By the tau-function, we actually mean its logarithm,
which only makes sense for the quasiclassical or Whitham-type
integrable hierarchies. Namely, the key properties are
the specific ``exchange'' relations (\ref{elem3}) which follow
from the Hadamard variational formula for the Green
function and the harmonic measure.
They have the form of integrability conditions
and thus ensure the existence of the tau-function.
The tau-function
corresponds to a particular solution of the universal Whitham
hierarchy \cite{KriW} and generalizes
the dispersionless tau-function which describes
deformations of simply-connected domains.

The algebro-geometric data associated
with the multiply-connected geometry include a Riemann
surface with antiholomorphic  involution,
the Schottky double of
the domain ${\sf D^c}={\bf C}\setminus {\sf D}$
endowed with particular holomorphic coordinates $z$ and
${\bar z}$ on the two sheets of the double, respecting the
involution. This Riemann surface has
genus $g = \#\{{\rm holes}\} -1$.
The (logarithm of) tau-function, ${\cal F}$,
describes small deformations of these data as functions
of an infinite set of independent deformation parameters
which are basically
harmonic moments of the domain.
These variables can be equivalently redefined as periods
of the generating one-form $S(z)dz$ over
non-trivial cycles on the double and the
residues of the one-forms $A_k(z)S(z)dz$, where
$A_k(z)\ \stackreb{z\to\infty}{\simeq}\ z^{-k}$ is some proper global
basis (\ref{a}) of harmonic functions.

We have obtained simple expressions for the period matrix,
the Abel map and the prime form on the Schottky double
in terms of the function ${\cal F}$. Specifically, all these
objects are expressed through second order
derivatives of the ${\cal F}$ in its independent variables.

The generalized dispersionless Hirota equations
on ${\cal F}$
for the multiply-connected case
(equivalent to the Whitham hierarchy)
are
obtained by incorporating the above mentioned
expressions
into the Fay identities.
As a result, one comes to a series
of quite non-trivial equations for
(second derivatives of) the function $F$,
which have not been written before (except for
certain relations for the second derivatives of
the Seiberg-Witten prepotential \cite{GMMM}).
When the Riemann surface degenerates
to the Riemann sphere with two marked
points, they
turn into Hirota equations of the dispersionless Toda
hierarchy.

Algebraic orbits of the universal Whitham hierarchy
describe the class
of domains which can be obtained as conformal images of
a ``half'' of a complex  algebraic curve with the
antiholomorphic involution under conformal maps
given by rational functions on the curve.
In particular, all domains having only a finite
amount of non-vanishing harmonic moments,
are in this class.
In this case one can define the curve by
a polynomial equation written explicitly in \cite{KM}.
This situation is
an analog of the (Laurent) polynomial conformal maps
in the simply-connected case and
literally corresponds to multi-support
solutions of matrix models
with polynomial potentials.
The definition of the tau-function for
multiply-connected domains proposed above
holds in a broader set-up of general algebraic
domains. It does not
rely on the finitness of the amount of non-vanishing
moments. In general any effective way to
describe the complex curve associated to
a multiply-connected domain by a system
of polynomial or algebraic equations is not known.
The curve may be thought of as a spectral curve
corresponding to a
generic finite-gap solution to the Toda lattice hierarchy.

\section*{Acknowledgments}

We are indebted to V.Ka\-za\-kov,
M.Mi\-ne\-ev-\-Wein\-stein, S.Na\-tan\-zon,
L.Takh\-ta\-jan and P.Wieg\-mann for
illuminating discussions and
especially to A.Levin for
very important comments on sect.~\ref{ss:eqs}.
We are also grateful to the referee
for a very careful reading of the manuscript
and valuable remarks.
%This work was supported in part by the Federal Program
%of the Russian Ministry of Industry, Science and Technology
%40.052.1.1.1112.
The work was also partialy supported by
NSF grant DMS-01-04621 (I.K.), RFBR under
the grant 01-01-00539 (A.M. and A.Z.),
by INTAS under the
grants 00-00561 (A.M.), 99-0590 (A.Z.)
and by the Program of support of scientific schools
under the grants 1578.2003.2 (A.M.), 1999.2003.2 (A.Z.).
The work of A.Z. was also partially supported by
the NATO grant PST.CLG.978817.
A.M. is grateful for the hospitality to the Max Planck
Institute of Mathematics in Bonn,
where essential part of this work has been done.

\renewcommand{\theequation}{A.\arabic{equation}}
\setcounter{equation}0
\section*{Appendix A. Theta functions and Fay identities}

Here we present some definitions and useful formulas from \cite{Fay}.
The Riemann theta function $\theta (\bW) \equiv \theta (\bW | T)$
is defined as
\be
\label{thetadef}
\theta (\bW) = \sum_{{\bf n}\in \bZ^g} e^{i\pi\bn\cdot T\cdot\bn +
2\pi i \bn\cdot\bW}
\ee
The theta function with
a (half-integer) characteristics $\bdelta =
(\bdelta_1, \bdelta_2)$, where $\delta_{\alpha} =
T_{\alpha \beta}\delta_{1,\beta}
+ \delta_{2,\alpha}$ and
$\bdelta_1, \bdelta_2 \in \frac{1}{2}\bZ ^{g}$ is
\be
\label{thetad}
\theta_{\bdelta} (\bW) = e^{i\pi\bdelta_1\cdot T\cdot\bdelta_1 +
2\pi i \bdelta_1\cdot(\bW+\bdelta_2)}
\theta (\bW + \bdelta) =
\sum_{{\bf n}\in \bZ^g} e^{i\pi(\bn+\bdelta_1)\cdot T\cdot(\bn+\bdelta_1) +
2\pi i (\bn+\bdelta_1)\cdot(\bW+\bdelta_2)}
\ee
Under shifts by a period of the lattice,
it transforms according to
\be
\label{thetaper}
\theta_\bdelta ({\bf W} + {\bf e}_\alpha) = e^{2\pi i \delta_{1,\alpha}}
\theta_\bdelta ({\bf W})
\\
\theta_\bdelta ({\bf W} + T_{\alpha\beta}{\bf e}_\beta) =
e^{- 2\pi i \delta_{2,\alpha} - i\pi T_{\alpha\alpha}
-2\pi iW_\alpha}\theta_\bdelta ({\bf W})
\ee
The prime form $E(z,\zeta)$ is defined as
\be
\label{primed}
E(z,\zeta ) = {\theta_\ast (\bW(z) -\bW(\zeta))\over
\sqrt{\sum_\alpha \theta_{\ast,\alpha}dW_\alpha(z)}
\sqrt{\sum_\beta \theta_{\ast,\beta}dW_\beta(\zeta)}}
\ee
where $\theta_\ast$ is any odd theta
function, i.e., the theta function
with {\em any odd} characteristic
$\bdelta^\ast$ (the characteristics is odd if
$4 \bdelta_1^\ast \cdot \bdelta_2^\ast = {\rm odd}$).
The prime form does not depend on the particular
choice of the odd characteristics.
In the denominator,
$$
\left. \theta_{\ast , \alpha}=
\theta_{\ast , \alpha}(0)=
\frac{\p \theta_{\ast}
(\bW )}{\p W_{\alpha}}\right |_{\bW =0}
$$
is the set of $\theta$-constants.

The data we use in the main text contain also a distinguished
coordinates on a Riemann surface: the holomorphic
co-ordinates $z$ and $\bar z$ on two different sheets, and we do not
distinguish, unless it is nessesary between the prime form (\ref{primed})
and a {\em function} $E(z,\zeta)\equiv E(z,\zeta)(dz)^{1/2}(d\zeta)^{1/2}$
``normalized" onto the differentials of distinguished co-ordinate.

Let us now list the Fay identities \cite{Fay} used
in the paper.
The basic one is Fay's trisecant formula
(equation (45) from p.~34 of \cite{Fay})
\be
\label{fay1}
\theta (\bW_1-\bW_3-\bZ)\ \theta (\bW_2-\bW_4-\bZ)\ E(z_1,z_4)E(z_3,z_2)
\\ + \,\,
\theta (\bW_1-\bW_4-\bZ)\ \theta (\bW_2-\bW_3-\bZ)\ E(z_1,z_3)E(z_2,z_4)
\\ = \,\,
\theta (\bW_1+\bW_2-\bW_3-\bW_4-\bZ)\ \theta (\bZ)\ E(z_1,z_2)E(z_3,z_4)
\ee
Here $\bW_i \equiv \bW (z_i)$. This identity
holds for {\em any} four points $z_1,\dots,z_4$ on a Riemann surface
and {\em any} vector $\bZ \in {\bf Jac}$.
In the limit
$z_3\to z_4\equiv\infty$ one gets
(formula (38) from p.~25 of \cite{Fay})
\be
\label{fay2}
{\theta ( \int_{\infty}^{z_1}d\bW +\int_{\infty}^{z_2}d\bW-\bZ)\
\theta (\bZ)\over
\theta (\int_{\infty}^{z_1}d\bW-\bZ)\
\theta (\int_{\infty}^{z_2}d\bW-\bZ)}
{E(z_1,z_2)\over E(z_1,\infty)E(z_2,\infty)} \\ = \,\,
d\Omega^{(z_1 , z_2)}(\infty) + \sum_{\alpha=1}^gdW_\alpha(\infty)
\,\, \p_{Z_{\alpha}}\!
\log\ {\theta (\int_{\infty}^{z_1}d\bW-\bZ)
\over\theta (\int_{\infty}^{z_2}d\bW-\bZ)}
\ee
where
\be
\label{abel3}
d\Omega^{(z_1 , z_2)}(\infty)=
d_z\log\ {E(z,z_1)\over E(z,z_2)}
\ee
is the normalized Abelian differential of the third kind
with simple poles at $z_1$ and $z_2$ and residues $\pm 1$.

Another relation from \cite{Fay} we use
(see e.g. (29) on p.20 and (39) on p.26) is
\be
\label{fay3}
{\theta (\bW_1-\bW_2-\bZ)\theta (\bW_1-\bW_2 + \bZ)\over
\theta ^2 (\bZ) E^2(z_1,z_2)} = \omega (z_1,z_2) +
\sum_{\alpha , \beta =1}^{g}
\left(\log\theta(\bZ)\right)_{,\alpha\beta}
dW_\alpha (z_1)dW_\beta (z_2)
\ee
where
$$
(\log \theta (\bZ))_{, \, \alpha \beta}=
\frac{\p^2 \log \theta (\bZ )}{\p Z_{\alpha}\p Z_{\beta}}
$$
and
\be
\label{bidif}
\omega (z_1,z_2) = d_{z_1}d_{z_2}\log E(z_1,z_2)
\ee
is the canonical bi-differential of the second kind
with the double pole at $z_1 =z_2$.

\renewcommand{\theequation}{B.\arabic{equation}}
\setcounter{equation}0
\section*{Appendix B. Degenerate Schottky double}

For an illustrative purpose we would like to
adopt some of the above formulas to the simplest
possible case, which is the Riemann sphere realized as
the Schottky double of the complement to the disk
of radius $r$. In this case
\be
\label{dWc}
dW = {dz\over z} = id\varphi = - {d{\bar z}\over\bar z}
\ee
is purely imaginary on the circle and
obviously satisfies the condition
$dW(z) + \overline{dW(\bar z)} = 0$.
Further, (cf. (\ref{n71}))
\be
\label{Wzc}
W(z) = \int_{\xi_0}^z dW = \int_{r}^z dW = \log {z\over r}
\ee
and
\be
\label{comapc}
w(z) = e^{W(z)} = {z\over r}
\ee
which is nothing but the
conformal map of the exterior of the circle
$|z|\geq r$ onto exterior of the unit circle $|w|\geq 1$.
Note that
on the ``lower" sheet of the double
\be
\label{Wzcb}
W(\bar z) = \int_{r}^{\bar z} dW = \log {r\over \bar z}
\ee
and instead of (\ref{comapc}) one gets
\be
\label{comapcb}
w(\bar z) = {r\over \bar z}
\ee
which is the conformal map of the
exterior of the disk $|\bar z|\geq r$ on the lower
sheet onto the {\em interior}
of the unit circle $|w|\leq 1$. The prime
form on the genus zero Riemann surface is
(cf. (\ref{primed}))
\be
\label{primec}
E(z_1,z_2) = {w_1-w_2\over\sqrt{dw_1}\sqrt{dw_2}}
\ee
where $w_i\equiv w(z_i)$.
Let us compute $E(\infty,\bar\infty)$, which
is understood in the main text, as
``normalized" on the values of the local
coordinates $z_\infty = 1/z$ and $\bar z_\infty =
1/\bar z$ in the points
$\infty$ and $\bar\infty$ on two sheets of the double. One gets
(cf. (\ref{Prime2}))
\be
\label{einfc}
E^2(\infty,\bar\infty) = - {(w(\infty)-w(\bar\infty))^2
\over (dw(\infty)/dz_\infty)(dw(\bar\infty)/dz_{\bar\infty})}
\ee
Substituting into (\ref{einfc}) the formulas
(\ref{comapc}), (\ref{comapcb}) and
\be
dw(\infty) = \lim_{z\to\infty} {dz\over r} = - \lim_{z\to\infty}
{z^2\over r}\ dz_\infty
\\
dw(\bar\infty) = r\ dz_{\bar\infty}
\ee
one finally gets
\be
E(\infty,\bar\infty)^2 = \lim_{z\to\infty} {z^2/r^2\over (z^2/r)r} =
{1\over r^2}
\ee
and this demonstrates that (\ref{corad}) indeed turns
into (\ref{sec7}) in the limit.

\end{document}